\documentclass[twocolumn,aps,prd,preprintnumbers,superscriptaddress,nofootinbib,10pt]{revtex4-1}
\usepackage{epsfig}
\usepackage{graphicx}
\usepackage{psfrag}
\usepackage{amsmath,amssymb}
\usepackage{colordvi}
\usepackage{amsfonts}
\usepackage{enumerate}
\usepackage{slashed}
\usepackage{color}
\usepackage{xcolor}

\usepackage{subfigure}

\usepackage{diagbox}

\newcommand{\bs}[1]{\boldsymbol{#1}}

\begin{document}
\title{Perturbative Calculations of Gravitational Form Factors at Large Momentum Transfer }

\author{Xuan-Bo Tong}
\affiliation{School of Science and Engineering, The Chinese University of Hong Kong, Shenzhen, Shenzhen, Guangdong, 518172, P.R. China}
\affiliation{University of Science and Technology of China, Hefei, Anhui, 230026, P.R.China}

\author{Jian-Ping Ma}
\affiliation{ CAS Key Laboratory of Theoretical Physics, Institute of Theoretical Physics, Chinese Academy of Sciences, Beijing 100190, China}

\author{Feng Yuan}
\affiliation{Nuclear Science Division, Lawrence Berkeley National
Laboratory, Berkeley, CA 94720, USA}

\begin{abstract}
We perform a perturbative QCD analysis of the gravitational form factors (GFFs) of nucleon at large momentum transfer. We derive the explicit factorization formula of the GFFs in terms of twist-3 and twist-4 light-cone distribution amplitudes of nucleon. Power behaviors for these GFFs are obtained from the leading order calculations. Numeric results of the quark and gluon contributions to various GFFs are presented with model assumptions for the distribution amplitudes in the literature. We also present the perturbative calculations of the scalar form factor $\langle P'|F^2| P\rangle$ for pion and proton at large momentum transfer. 
\end{abstract}
\maketitle
      
\section{Introduction}
The physics of the gravitational form factors (GFFs) has attracted a great interest of hadron physics community in recent years. The experimental investigations of the GFFs are one of the main goals for the ongoing JLab 12 GeV program~\cite{Dudek:2012vr} and the future electron-ion colliders~\cite{Accardi:2012qut,AbdulKhalek:2021gbh,Anderle:2021wcy}. The GFFs are defined as the form factors of the energy-momentum tensor (EMT), which describe the elastic interaction between the graviton and the particles~\cite{Kobzarev:1962wt,Pagels:1966zza,Ji:1996ek}. 

The EMT of a field theory in the Minkowski space is  derived from the space-time translation invariance of the action by the Noether theorem~\cite{Noether:1918zz}. The EMT obtained in this way is called the canonical EMT and normally not symmetric with respect to its Lorentz indices. However, the physical concept of the GFFs actually stems from another kind of the EMT which plays a unique role in the Einstein theory of gravity~\cite{Einstein:1916vd}:
 \begin{align}
T^{\mu\nu}(x) =\frac{2}{\sqrt{-g}} \frac{\delta S_M}{\delta g_{\mu\nu}(x)} \ ,
\end{align}
where $g_{\mu\nu}(x)$ is the classical gravitational field in the general relativity and $S_M$ is the matter action of the system.  This form of EMT is automatically symmetrized and differs from the canonical one by a total derivative, hence it is called symmetric EMT or Belinfante-improved EMT~\cite{BELINFANTE1939887,belinfante1940current}.

To explain the implication of GFFs, we can consider a hadron system in the classical gravitational background.
When the curvature of the gravitational field is weak enough, we can expand the metric field around the Minkowski metric $\eta_{\mu\nu}=(1,-1,-1,-1)$,
\begin{align}
g_{\mu\nu}(x)=\eta_{\mu\nu}+h_{\mu\nu}(x)\ .
\end{align}
Then what we obtain is a flat-space action in which the perturbation field $h_{\mu\nu}(x)$ is equivalent to a spin-2 field and the corresponding particle can be considered as the graviton. In the action, it is interesting to find that the graviton couples with the EMT operator in the following form:
\begin{align}
\int d^4x ~ h_{\mu\nu }(x) T^{\mu\nu }_{QCD}(x)~.
 \end{align}
That means that when a hadron $N$ scatters with the graviton elastically, the only way the quarks and gluons inside the hadron can interact with them is through EMT operator. Therefore, this scattering is describe by the following transition matrix 
\begin{align}
\langle N(P') |T^{\mu\nu}_{QCD}| N(P) \rangle\ .
\end{align}
In general, this matrix is hard to evaluate due to the color confinement of QCD. The conventional approach is to apply the Lorentz symmetry on the EMT matrix and parametrize it into the so-called form factors, i.e GFFs~\cite{Kobzarev:1962wt,Pagels:1966zza,Ji:1996ek}, which only depend on the momentum transfer squared $t=(P'-P)^2$. Different form factors correspond to different Lorentz structures and hence encode different physical information of the hadron structure, such as mass~\cite{Ji:1994av,Ji:1995sv,Ji:2021mtz,Ji:2021qgo,Lorce:2017xzd,Hatta:2018sqd,Tanaka:2018nae,Metz:2020vxd,Lorce:2021xku,Yang:2018nqn,He:2021bof}, spin~\cite{Jaffe:1989jz,Ji:1996ek,Leader:2013jra,Ji:2020ena}, and mechanical properties~\cite{Polyakov:2002yz,Polyakov:2018zvc,Shanahan:2018nnv,Lorce:2018egm,Varma:2020crx,Panteleeva:2021iip,Freese:2021czn,Burkert:2018bqq,Kumericki:2019ddg,Dutrieux:2021nlz,Burkert:2021ith}, and gravitational multipoles~\cite{Ji:2021mfb}. The physical content from quarks and gluons can be also explored based on the gauge-invariant separation of the total EMT $T_{QCD}^{\mu\nu}$ into the quark part $T^{\mu\nu}_q$ and the gluon part $T^{\mu\nu}_g$, and the associated GFFs can be defined accordingly~\cite{Ji:1996ek}. 

Although it is not realistic to probe the GFFs directly through the graviton-hadron scattering in experiments, they can be alternatively extracted from the generalized parton distributions (GPDs) in the hard exclusive processes~\cite{Ji:1996ek,Ji:1996nm,Mueller:1998fv,Diehl:2003ny,Belitsky:2005qn} 
, e.g., from the deeply virtual Compton scattering~\cite{Ji:1996ek,Ji:1996nm,Radyushkin:1996nd,dHose:2016mda,Kumericki:2016ehc} and the deeply virtual meson production~\cite{Collins:1996fb,Mankiewicz:1997bk,Favart:2015umi}. Meanwhile, they can also be explored in the time-like processes such as the hadron productions in the two-photon collisions, see, for example, a recent analysis~\cite{Kumano:2017lhr} on the experimental data from the Belle Collaboration at KEK~\cite{Belle:2015oin}. 

In particular, the explorations of the quark GFFs in the limited momentum-transfer region from experiments have been reported in~\cite{Burkert:2018bqq,Kumericki:2019ddg,Dutrieux:2021nlz,Burkert:2021ith}.
Meanwhile, there has been great research interest of the near-threshold heavy quarkonium photo-production and its potential contribution to constrain the gluonic GFFs~\cite{Kharzeev:1995ij,Kharzeev:1998bz,Hatta:2018ina,Kharzeev:2021qkd,Hatta:2019lxo,Boussarie:2020vmu,Hatta:2021can,Guo:2021ibg,Sun:2021gmi,Sun:2021pyw,Mamo:2021tzd,Mamo:2022eui,Wang:2021dis,Kou:2021qdc,Kou:2021bez}.
These discussions are essential to understand the nucleon mass decomposition
~\cite{Ji:1994av,Ji:1995sv,Ji:2021mtz,Ji:2021qgo,Lorce:2017xzd,Hatta:2018sqd,Metz:2020vxd,Lorce:2021xku,Yang:2018nqn,He:2021bof}, the pressure and shears forces inside the hadron system~\cite{Polyakov:2002yz,Polyakov:2018zvc,Shanahan:2018nnv,Lorce:2018egm,Varma:2020crx,Panteleeva:2021iip,Freese:2021czn,Burkert:2018bqq,Kumericki:2019ddg,Dutrieux:2021nlz,Burkert:2021ith} as well as the momentum -current gravitational multipoles of hadrons~\cite{Ji:2021mfb}.

From the theory side, the GFFs can be computed non-perturbatively in the framework of the lattice QCD, see, e.g., the pioneer works in Refs.~\cite{Hagler:2007xi,Deka:2013zha,Shanahan:2018nnv,Shanahan:2018pib}. On the other hand, if the momentum transfer $(-t)$ is sufficiently large, one can also carry out the perturbative analysis thanks to the asymptotic freedom of the QCD. This is the main subject of this paper. We will perform a systematic analysis on the complete set of the nucleon GFFs at large momentum transfer, including both quark and gluon sectors. The methodology is based on the widely-used Efremov-Radyushkin-Brodsky-Lepage (ERBL) formalism in hard exclusive processes~\cite{Lepage:1979za,Brodsky:1981kj,Efremov:1979qk,Chernyak:1977as,Chernyak:1980dj,Chernyak:1983ej,Belitsky:2002kj} and the progresses in the classification of the nucleon light-front wave function involving the orbital angular momentum~\cite{Ji:2002xn,Ji:2003yj,Belitsky:2002kj} as well as the power counting rule~\cite{Brodsky:1973kr,Matveev:1973ra,Ji:2003fw}.  We will mainly focus on the nucleon GFFs and derive their factorization formulas in terms of twist-3 or twist-4 nucleon light-cone distribution amplitudes~\cite{Braun:1999te,Braun:2000kw}. Parts of our study on the gluonic GFFs as well as the pion GFFs have been reported in the letter recently~\cite{Tong:2021ctu}. 

The rest of the paper is organized as follows. In Sec.~II, an introduction of the quark and gluon EMT and the GFFs will be given. In Sec.~III, we perform the perturbative analysis of the nucleon GFFs at large momentum transfer. 
In Sec.~IV, we will present the numeric evaluation of the individual GFFs based on model assumptions for the twist-3 and twist-4 distribution amplitudes. In Sec.~V, the scalar form factors at large $(-t)$ will be discussed for both pion and proton. Sec.~VI is the conclusion of the paper.

 \section{Gravitational form factors}

In this section, we will first present the explicit expressions of the symmetric QCD EMT and discuss some important features of it.
\par
\subsection{Energy-Momentum Tensor}
 The symmetric QCD energy momentum tensor is defined as
\begin{align}
T^{\mu\nu }_{QCD}= T^{\mu\nu}_q+T^{\mu\nu}_g \ .
\label{eq_totalEMT}
\end{align}
The quark and gluon sectors are given by
\begin{align}
T^{\mu\nu}_q&=\frac{1}{4} \bar \psi_f \left(-i \overleftarrow D^\mu  \gamma^\nu-i \overleftarrow D^\nu  \gamma^\mu +i \overrightarrow D^\mu  \gamma^\nu+i \overrightarrow D^\nu  \gamma^\mu \right)\psi_f~,
\notag \\
T^{\mu\nu}_g&=F^{a,\mu\lambda }F_{\lambda}^{a,\nu}+\frac{1}{4}g^{\mu\nu }F^{a,\sigma \rho }F^a_{\sigma \rho }~,
\label{eq_EMT}
\end{align}
where $\psi_f$ is a quark field of flavor $f$, and $F^a_{\mu\nu}=\partial_\mu A^a_\nu-\partial^\nu A^a_\mu-g_s f^{abc} A^b_\mu A^c_\nu
$ is the strength tensor of the gluon field $A^{a,\mu}$, and the covariant derivatives are defined as
\begin{align}
\overrightarrow D_\mu =\overrightarrow \partial_\mu +i g_s A^a_\mu T^a,\quad
\overleftarrow D_\mu =\overleftarrow \partial_\mu -i g_s A^a_\mu T^a \ .
\end{align}
The sum of the quark flavors and colors are implied.

There are several features that we would like to emphasize. First, the total EMT in Eq.~(\ref{eq_totalEMT}) is conserved due to the space-time translation invariance, i.e.  $\partial_\nu T^{\mu\nu }_{QCD} =0$. However, the gluon or quark part of the EMT is not conserved individually. Instead, the derivative of the EMT operators are related to the twist-four operators~\cite{Braun:2004vf,Tanaka:2018wea},
\begin{align}
\partial_\nu T_q^{\mu\nu}&=\bar  \psi_f g F^{\mu\nu} \gamma_\nu \psi_f~
\label{eq:EMTcon1}
,\\
\partial_\nu T_g^{\mu\nu}&=F_{\nu}^{\ \mu} D_\alpha F^{\alpha\nu} \ .
\label{eq:EMTcon2}
\end{align}
The above equations are derived by applying the equations of motion for the quark and gluon fields: $(i\stackrel{\rightarrow}{\slashed D} -m_f)\psi_f=0$ and $\partial^\mu F^a_{\mu\nu} -g f^{abc} A^{b\mu} F^c_{\mu\nu} =g \bar \psi \gamma_\nu T^a \psi$. With the EOM of the gluon strength tensor, one can also find the right sides of Eqs.(\ref{eq:EMTcon1}) and (\ref{eq:EMTcon2}) cancel out each other. This confirms the conservation of the total EMT operator explicitly. Therefore, the total EMT operator is UV finite and scale-independent~\cite{Freedman:1974gs,Nielsen:1975ph,Nielsen:1977sy,Collins:1976yq} while the quark or gluon sector of the EMT is divergent and needs regularization and renormalization~\cite{Hatta:2018sqd,Tanaka:2018nae}. The renormalization of the EMT is closely related to the trace of this operator. Classically, the quark or gluon EMT is  traceless if the quark masses are neglected:
\begin{align}
g_{\mu\nu}  T^{\mu\nu}_{q,g}=0~,
\end{align}
which is also the consequence of the conformal symmetry in the massless QCD. This feature will be broken at the quantum loop level, called trace anomaly~\cite{Freedman:1974gs,Nielsen:1975ph,Nielsen:1977sy,Collins:1976yq,Tarrach:1981bi,Ji:1995sv}:
\begin{align} 
T^\mu_\mu=\frac{\beta(g_s)}{2 g_s} F^{a,\mu\nu} F^a_{\mu\nu}~,
\label{eq_traceanomaly}
\end{align}
where $\beta(g_s)$ is the QCD beta function and the terms $(1+\gamma_m) m_f\bar \psi_f \psi_f$ should be added in the massive case. 
\par 
Second, in the covariant quantization of QCD, the EMT operators in Eqs.~(\ref{eq_totalEMT},\ref{eq_EMT}) are not completed rigorously. For example, in the Faddeev-Popov quantization procedure for the gauge theory one has to introduce the ghost and gauge-fixing terms in the Lagrangian, thus there are corresponding terms in the QCD EMT~\cite{Freedman:1974gs}. However, these terms are ensured to be absent in the physical matrix elements by the BRST symmetry~\cite{Joglekar:1975nu} and hence have no observable effects. Therefore, we have excluded them in our study of the GFFs.

\subsection{Nucleon Gravitational Form Factors}
Nucleon is a spin-$1/2$ particle, and the quark or gluon gravitational form factors can be parameterized as \cite{Ji:1996ek}:
\begin{align}
&\langle P',s' | T^{\mu\nu}_a(0)|P,s\rangle
= \bar u_{s}(P') \Bigg[ 
A_a(t) \gamma^{(\mu} \bar P^{\nu)}
\notag \\
& +B_a(t)\frac{i \bar P^{(\mu }\sigma^{\nu  )\rho}
 \Delta_\rho }{2 M}
+ C_a(t) \frac{\Delta^\mu \Delta^\nu -g^{\mu\nu} \Delta^2}{M}
 \notag \\
&
+\overline C_a(t) M g^{\mu\nu}
 \Bigg] u_s(P)\ ,\label{eq_GFF1}
\end{align}
where $a=q,g$ and the brace for the Lorentz indices $\mu,\nu$ denote the symmetrization known as $A^{(\mu} B^{\nu )}\equiv(A^\mu B^\nu+A^\nu B^\mu)/2$. $P$ and $P'$ are the momenta of the initial and final particle respectively, satisfying the on-shell condition $P^2={P'}^2=M^2$, where $M$ is the nucleon mass. $\bar P$ is the average momentum, $\Delta$ is the momentum transfer and $t$ is the momentum transfer squared: 
\begin{align}
\bar P=(P+P')/2,\quad \Delta=P'-P,\quad t=\Delta^2\ .
\end{align}
The nucleon state is covariant normalized, $\langle P'|P\rangle=2 P^0(2\pi)^3\delta^3(\vec P'- \vec P)$.  $u_s(P)$ is the Dirac spinor of nucleon normalized as $\bar u_{s'}(P)u_s(P)=2M\delta_{s s'}$. 
One can utilize the well-known Gorden identity for the spinor,
\begin{align}
2M \bar u_{s'}(P') \gamma^\alpha u_s(P)=\bar u_{s'}(P')(2 \bar P^\alpha +i\sigma^{\alpha \kappa} \Delta_\kappa)  u_s(P)~,
\end{align}
to obtain another parametrization for the GFFs:
\begin{align}
&\langle P',s' | T^{\mu\nu}_a(0)|P,s\rangle
=\bar u_{s'}(P') \Bigg[2 J_a(t)\gamma^{(\mu} \bar P^{\nu)}
\notag \\&
-B_a(t) \frac{\bar P^{\mu} \bar P^{\nu}}{M} 
+ C_a(t) \frac{\Delta^\mu \Delta^\nu -g^{\mu\nu} \Delta^2}{M}
\notag \\& 
+\overline C_a(t) M g^{\mu\nu}
 \Bigg] u_s(P)\ ,\label{eq_GFF2}
\end{align}
where $J_a(t)=[A_a(t)+B_a(t)]/2$. Here we follow the notations in Refs.~\cite{Ji:1996ek,Ji:1996nm} for the $C$ form factors. They have also been referred as $D$ or $d_1$ form factors in Refs.~\cite{Polyakov:2002yz,Polyakov:2018zvc,Burkert:2018bqq,Shanahan:2018nnv,Kumericki:2019ddg} with different normalizations: $D(t) = 4/5 d_1(t) = 4 C(t)$.
\par 
From the above expressions, one can observe the quark and gluon GFFs are the Lorentz invariants of the momentum transfer squared $t$. As shown in Eqs.~(\ref{eq:EMTcon1},\ref{eq:EMTcon2}), the quark and gluon parts of EMT are not conserved and require renormalization, and the corresponding GFFs should also depend on the renormalization scale~\cite{Hatta:2018sqd,Tanaka:2018nae}. For each form factor, we can also define the total GFF, e.g. $ C(t)=\sum_{a=q,g}C_a(t)$. They are, on the other hand, renormalization scale independent due to the conservation of the total EMT. This feature has further implication on the GFFs: 
\begin{align}
\langle P' |  \partial_\nu  T_{QCD}^{\mu\nu }  | P\rangle
=& \langle P' | i\Big[ \hat P_\nu, T_{QCD}^{\mu\nu } \Big] | P\rangle
\notag\\
=& 
i\Delta_\nu \langle P' |T_{QCD}^{\mu\nu}| P\rangle 
=0\ ,
\end{align}
where the Hesienberg equation of the EMT operator is applied. The last equality yields 
\begin{align}
\Delta_\nu \langle P',s' |T_{QCD}^{\mu\nu}| P,s\rangle 
= \overline C(t)  M \Delta^\mu \bar u_{s'}(P') u_{s}(P)=0\ ,
\end{align}
where the terms related to the $A,B,C$-form factors vanish automatically from its own tensor structure or the on-shell condition of the Dirac spinor, $\slashed P u(P)=M u(P)$. Therefore, an important constraint on the $\overline { C}$-GFFs is obtained:
\begin{align}
\overline C(t)=\sum_{a=q,g} \overline  C_a(t)=0 \ .
\label{eq:cbarvanish}
\end{align}
This constraint works regardless of the value of the momentum transfer squared $t$ and will serve as an important consistent check on our perturbative calculations of the quark and gluon GFFs at large momentum transfer. The details will be presented in Sec.~\ref{sec:cbarcheck}.
\par 
The GFFs $\overline C_{q,g}(t)$ play an important role in the nucleon mass reconstruction since it is closely related to the trace anomaly $F^2$ and its value at $t=0$ determines the contributions of the so-called quantum anomalous energy~\cite{Ji:1994av,Ji:1995sv,Ji:2021mtz,Ji:2021qgo,Hatta:2018sqd,Boussarie:2020vmu}. 
The $C_{q,g}(t)$ form factors in Eqs.~(\ref{eq_GFF1},\ref{eq_GFF2}) are proposed to describe the mechanical properties such as pressure and shear forces distribution inside the nucleon~\cite{Polyakov:2002yz,Polyakov:2018zvc,Panteleeva:2021iip,Freese:2021czn}. This interpretation uniquely requires the whole knowledge of $t$-dependence from $C$-form factor. 
\par 
In addition, the $A_{q,g}$ form factor at zero momentum transfer can be interpreted as the longitudinal momentum fraction that the parton carries inside the nucleon in the infinite momentum frame with $\sum_{a=q,g}A_a(0)=1$. The $J$-form factor $J_a(t)=[A_a(t)+B_a(t)]/2$ at $t=0$ describe the partitions of the quark and gluon angular momentum from the nucleon spin satisfying the constraint $\sum_{a=q,g} J_a(0)=\frac{1}{2}$, which is known as the Ji's sum rule in the literature~\cite{Ji:1996ek}.

In the following sections, we will carry out a systematic investigation on the nucleon GFFs for the quarks or gluons at large $(-t)$. 

\section{Perturbative Analysis at large momentum transfer\label{sec:perturbative}}
 
In general, the nucleon GFFs are non-perturbative due to the color confinement.
Nonetheless, in the large momentum transfer limit, there are indeed perturbative calculable effects in the GFFs. The physics behind is that the highly virtual graviton has short enough wavelength to resolve the structure of the nucleon by the short space-time interactions. The partons participated in this interaction are weakly coupled but endure with hard gluon exchanges to pass the large momentum transfer from the initial nucleon to the final nucleon. These effects can be calculated using perturbation theory due to the asymptotic freedom of QCD. On the other hand, the interactions among those active partons and the other spectator partons inside the nucleon are still of long range and hence strong. In this section, we will demonstrate that this short-range and long-range physics can be factorized at lowest order perturbation theory, leading to the factorization theorem of the GFFs in terms of twist-3 or twist-4 nucleon light-cone distribution amplitudes. Our derivations follow the ERBL formalism ~\cite{Lepage:1979za,Brodsky:1981kj,Efremov:1979qk,Chernyak:1977as,Chernyak:1980dj,Chernyak:1983ej,Belitsky:2002kj} with development on the nucleon light-front wave function involving zero or one unit of orbital angular momentum~\cite{Ji:2002xn,Ji:2003yj,Belitsky:2002kj}. 
 \par 
We will first introduce the helicity-amplitude with appropriate tensor projection to isolate each GFF, and choose a reference frame. Later, we will review the three-quark Fock state of the nucleon in the literature.
Based on these materials, we will carry out the perturbative analysis of the GFFs through the helicity-conserved and helicity-flip amplitudes, respectively. Finally, we summarize the factorization results and perform consistent checks.

\subsection{Helicity Amplitude of EMT \label{sec:helicityamp}}
Since the nucleon is a spin-1/2 particle, the parametrization of the EMT amplitude involves the Dirac structures as well as the tensor structures.  The corresponding form factors can be extracted from the nucleon helicity amplitude with appropriate tensor projections. In the high energy, we will find that the helicity-conserved amplitudes yield the $A$-form factors, and the helicity-flip amplitudes lead to other GFFs.
\par 
As presented in Eq.~(\ref{eq_GFF2}), there are two Dirac structures involved in the EMT amplitude: $\bar u_{s'} \gamma^{\mu} u_{s}$ and $\bar u_{s'}u_{s}$, corresponding to the chiral-odd and chiral-even Dirac bilinear spinors. In the large momentum transfer limit, the nucleon mass can be neglected, and hence the chirality becomes a good quantum number to identify the nucleon state, i.e.,
\begin{align}
\gamma_5 u_{\uparrow/\downarrow}(P)=\pm u_{\uparrow/\downarrow}(P)~,
\end{align}
where $u_{\uparrow/\downarrow}(P)$ is the chirality eigenstates, named as right- or left-handed spinors. Besides, for a massless spin-1/2 particle, its chirality is known to be equivalent to its helicity, where helicity is defined by the projection of the spin along the three-momentum direction. We will use helicity as the terminology in the following paragraphs.  Now we choose the appropriate helicity eigenstates for the initial and final nucleons along with tensor projections to extract the GFFs. 
\paragraph*{Helicity-conserved Amplitude}  $A$-form factor can be extracted form the helicity-conserved amplitude:
\begin{align}
&\langle P'_\uparrow | T^{\mu\nu}_a|P_\uparrow\rangle
=A_a(t)\bar u_\uparrow(P') \gamma^{(\mu} \bar P^{\nu)} 
u_\uparrow(P)\ ,
\label{eq_consAmp}
\end{align}
where $B$-form factor are power-suppressed as shown later and hence neglected in the above equation. Since only $A$-form factor is relevant at the leading power, we do not need to assign the tensor projection. Here we used the amplitude with positive helicity. The amplitude for the negative helicity can also be used and leads to the same result.
\paragraph*{Helicity-flip Amplitude} On the other hand, helicity-flip amplitude can be used to extract the GFFs ${ B}_a$, ${ C}_a$ and $\overline{ C}_a$:
\begin{align}
&\langle P'_\uparrow| T^{\mu\nu}_a|P_\downarrow\rangle
=\bar u_{\uparrow}(P') \left[
-\frac{B_a(t) }{M}\bar P^{\mu} \bar P^{\nu} \right.
\notag \\
&
\left.+\frac{C_a(t)}{M}  (\Delta^\mu \Delta^\nu -g^{\mu\nu} \Delta^2)
+\overline C_a(t)M g^{\mu\nu}
 \right] u_\downarrow(P)\ .
 \label{eq_flipAmp}
\end{align}
To further isolate the GFFs, we can define the following projection tensors,
\begin{eqnarray}
\Gamma_1^{\mu\nu}&=&4\bar P^\mu\bar P^\nu-\bar P^2 g^{\mu\nu} \ ,\\
\Gamma_2^{\mu\nu}&=&\Delta^\mu\Delta^\nu-\Delta^2 g^{\mu\nu}/4 \ .
\end{eqnarray}
Notice that $g^{\mu\nu}\Gamma_{1,2}^{\mu\nu}=0$ in the massless limit. Applying the above, we can derive
\begin{eqnarray}
{ B}_a(t)&=&\frac{2M\left(\Gamma_2^{\mu\nu}-3\Gamma_1^{\mu\nu}\right)\langle P'_\uparrow| T^{\mu\nu}_a|P_\downarrow\rangle}{(\Delta^2)^2\bar u_\uparrow(P')u_\downarrow(P)} \ ,\\
{ C}_a(t)&=&\frac{M\left(3\Gamma_2^{\mu\nu}-\Gamma_1^{\mu\nu}\right) \langle P'_\uparrow| T^{\mu\nu}_a|P_\downarrow\rangle}{2(\Delta^2)^2\bar u_\uparrow(P')u_\downarrow(P)} \ .
\label{eq_tensor}
\end{eqnarray}
From the traceless feature of the EMT in our calculations at this order, we can also derive $\overline { C}$ form factor:
\begin{align}
\overline C_a(t)=& \frac{-t}{16M^2 } \big [B_a(t)-12 C_a(t) \big].
\notag \\
=&\frac{\Gamma_2^{\mu\nu}\langle P'_\uparrow| T^{\mu\nu}_a|P_\downarrow\rangle}{M (\Delta^2)\bar u_\uparrow(P')u_\downarrow(P)} \ .
\end{align} 
\par \noindent
As mentioned above, due to the conservation of the EMT, 
the total $\overline C $-form factor should vanish. That renders
\begin{align}
\Gamma_2^{\mu\nu}\langle P'_\uparrow| T^{\mu\nu}_{QCD}|P_\downarrow\rangle=0 \ .
\end{align}
Furthermore, the total contributions to the $B$ and $C$ form factors follow
\begin{eqnarray}
B(t)&=&-\frac{6 M \Gamma_1^{\mu\nu}\langle P'_\uparrow| T^{\mu\nu}_{QCD}|P_\downarrow\rangle}{(\Delta^2)^2\bar u_\uparrow(P')u_\downarrow(P)} \ ,\\
C(t)&=&- \frac{M\Gamma_1^{\mu\nu}\langle P'_\uparrow| T^{\mu\nu}_{QCD}|P_\downarrow\rangle }{2(\Delta^2)^2\bar u_\uparrow(P')u_\downarrow(P)} \ . 
\end{eqnarray}
As a consequence, we find that there is a nontrivial relation between the total $B$- and $C$-form factors at large momentum transfer at the leading order perturbation theory:
\begin{align}
B(t)=12C(t)\ . \label{eq:bcrelation}
\end{align}
This relation along with the constraint $\overline C(t)=0$ will serve as the consistent checks of our calculations. 
\par 
\subsection{Breit Frame}
In order to facilitate the computation, we need to fix the reference frame explicitly. Since the GFFs are the Lorentz invariant, one can choose any frame in principle. Particularly, we use the so-called Breit frame in the calculation where the initial nucleon is along the $z$-axes with high energy and the final nucleon is along the opposite  direction:
\begin{align}
&P^\mu=(P^+,P^-,\bs P_\perp)
=\frac{1}{\sqrt{2}}\left(\sqrt{-t},\ 0,\ \bs 0_\perp \right),
\notag \\
&P'^\mu=(P'^+,P'^-,\bs P'_\perp)
=\frac{1}{\sqrt{2}}\left(0,\ \sqrt{-t},\ \bs 0_\perp \right),
\notag \\
&\Delta^\mu=(\Delta^+,\Delta^-,\bs \Delta_\perp)=\left(-\sqrt{-t},\ \sqrt{-t},\ \bs 0_\perp\right).
\end{align}
Here the large momentum transfer limit $-t\gg M^2$ is applied and the light-cone coordinates are used where $a^+=(a^0+a^3)/\sqrt{2}$, $a^-=(a^0-a^3)/\sqrt{2}$ and $a_\perp^2=-\bs a_\perp^2$. In the Breit frame,  the momenta of the initial state and final state nucleons as well as the momentum transfer contain only the longitudinal components. Hence, all the large scales in the partonic subprocesses must come from the longitudinal momentum and  make the power counting of the perturbative part more transparent. 
\subsection{Three-valence Quark Fock State of the Nucleon}
In the Breit frame, both the initial and final nucleons are traveling along the longitudinal direction and carrying large momentum. 
In this frame, one can apply the light-front Fock state expansion~\cite{Brodsky:1997de}. 
In general, there are infinite number of Fock states with light-front wave functions in this expansion. However, for an exclusive process at the large momentum transfer, one can conclude from the general power counting rule that the leading contributions come from the terms with minimal numbers of partons and the fewest orbital angular momentum (OAM) content~\cite{Brodsky:1973kr,Matveev:1973ra,Ji:2003fw}.  In the following, we will review the three-valence quark Fock state of the nucleon~\cite{Ji:2002xn,Ji:2003yj}.
 
\par
For the nucleon, the leading Fock state is made of  three valence quarks, i.e. $u,u,d$. Each quark can carries the helicity $\lambda_i=\pm 1/2$. 
The total helicity of the valence quarks has the value as $3/2,1/2,-1/2,-3/2$ and the corresponding Fock states can be denoted
\begin{align}
  \big| P_\uparrow  \big\rangle=\big| P_\uparrow \big\rangle^{l_z=2}_{-3/2}+\big| P_\uparrow \big\rangle^{l_z=1}_{-1/2}+\big| P_\uparrow \big\rangle^{l_z=0}_{1/2}+\big| P_\uparrow \big\rangle^{l_z=-1}_{3/2}.
 \end{align}
where the upper subscript represents the total quark helicity $\lambda=\sum_i \lambda_i$ and the superscript stands for the projection of OAM along the $z$ direction determined from the angular momentum sum rule $l_z=\Lambda -\lambda$. Similarly, the left-handed proton can be expressed as 
 \begin{align}
  \big| P_\downarrow  \big\rangle=\big| P_\downarrow \big\rangle^{l_z=1}_{-3/2}+\big| P_\downarrow \big\rangle^{l_z=0}_{-1/2}+\big| P_\downarrow \big\rangle^{l_z=-1}_{1/2}+\big| P_\downarrow \big\rangle^{l_z=2}_{3/2}\ .
 \end{align}
Therefore, the proton helicity amplitude of EMT
\begin{align}
 \langle P', \Lambda' | T^{\mu\nu}_a|P,\Lambda\rangle \ ,
 \end{align}
can be expressed with the above Fock state components. 

Then the nucleon EMT amplitudes are factorized into the partonic EMT amplitudes multiplied by the light-front wave function from the initial and final state nucleons. In the high energy scattering, the light quark mass can be neglected, and the parton helicity is conserved.  
That means the nucleon helicity-flip amplitude can only be obtained by involving a non-zero OAM from either initial or final state Fock components. 
In fact, this OAM along the longitudinal direction is achieved by the parton's intrinsic transverse motion and hence introduces some transverse momentum factors in the phase space integral which normally scale as $\Lambda_{QCD}$.  In the end, this factor will be compensated by the large momentum scale ($-t$ in our case) and lead to power suppression. Therefore, more OAM involved, more power suppression will be obtained~\cite{Ji:2003fw}.
\par For the nucleon helicity-conserved amplitude in Eq.(\ref{eq_consAmp}), we have
 \begin{align}
\langle P'_ { \uparrow }  |
  {T^{\mu\nu }_a} | P_ \uparrow  \rangle=\langle P'_ { \uparrow }  |^{l'_z=0}_{1/2} 
 {T^{\mu\nu }_a} | P_ \uparrow  \rangle^{l_z=0}_{1/2}\ ,
\end{align}
where the power suppressed contributions from $\Delta l_z=l'_z-l_z \neq 0$ are neglected. However, for the nucleon helicity flip amplitude in Eq.~(\ref{eq_flipAmp}), one unit of OAM is needed to achieve the spin flip. Hence, the leading contributions come from the components with $\Delta l_z=1$,
\begin{align}
\langle P'_{ \uparrow }  |
 {T_a^{\mu\nu}} |P_{\downarrow }\rangle 
=&\langle P'_{ \uparrow }  |^{l'_z=0}_{1/2}  
 {T_g^{\mu\nu}} | P_{\downarrow }\rangle ^{l_z=-1}_{1/2}
 \notag \\ 
&+ \langle P'_{ \uparrow }  |^{l'_z=1}_{-1/2}  
 {T_a^{\mu\nu}} | P_{ \downarrow }\rangle ^{l_z=0}_{-1/2}\ ,
\end{align}
where the helicity flip for the first term is from the initial state OAM and the second one from the final state OAM. From the parity and time reversal invariance symmetry, one can show that they in fact have the same contributions to the GFFs at large $-t$. Here we only focus on the first term.

\par 
In the followings, the expressions of the relevant Fock state components involved in our evaluation for the nucleon helicity amplitude are given~\cite{Ji:2002xn,Ji:2003yj}:
\begin{align}
 \big| P_\uparrow \big\rangle_{1/2}^{l_z=0}= & \int \
    \frac{[d x][d^2 \bs k]}{\sqrt{24 x_1 x_2 x_3}}
 \psi_1
   ( \kappa_1,\kappa_2,\kappa_3)        \notag
 \\ & \times
      \ \big |\{x_i P+ \bs  k_i \}\big \rangle_{1/2},
 \label{eq_fock1} \\ 
 \big| P_\downarrow \big\rangle_{1/2}^{l_z=-1}= & \int \
    \frac{[d x][d^2 \bs k]}{\sqrt{24 x_1 x_2 x_3}}
   \big[  k_{1L} \psi_3+  k_{2L} \psi_4 \big]
   ( \kappa_2,\kappa_1,\kappa_3) 
       \notag \\ & \times
      \ \big |\{x_i P+ \bs  k_i \}\big \rangle_{1/2}~,
       \label{eq_fock2} 
 \end{align}
 where the three-valence-quark state is defined by
 \begin{align}
\big|
\{p_i\} \big\rangle_{1/2}=&
\frac{\epsilon_{abc}}{\sqrt{6}}
  \hat u_{a,\uparrow}^\dagger (p_{ 1} ) 
   \big [\hat u_{b,\downarrow}^\dagger (p_{2 })
  \hat d_{c,\uparrow}^\dagger ( p_{3 })
      \notag \\ & \  - \hat d_{b,\downarrow}^\dagger( p_{2 })\hat  u_{c,\uparrow}^\dagger ( p_{3 })\big]\big|0 \big\rangle~.
         \label{eq_fock3}
\end{align}
In the above expressions, $\hat u^\dagger_{a,\lambda}$ and $\hat d^\dagger_{a,\lambda}$ are the creation operators of the $u,d$ quarks with the helicity $\lambda$ and color indices $a$ in the fundamental representation of SU(3) group, normalized as $\{\hat u_{a,\lambda}(k), \hat u^\dagger_{b,\lambda'}(k')\}=\delta_{\lambda \lambda'}\delta_{ab}2k^+\delta(k^+-k'^+)\delta^{(2)}(\bs k'-\bs k)$. $\psi_1,\psi_3,\psi_4$ are the light-front wave functions of the proton depending on $\kappa_i\equiv(x_i, \bs k_i)$: $x_i$ is the longitudinal momentum fraction of the parton with $\sum_i x_i=1$; $\bs k_i$ is the intrinsic transverse momentum of the parton with $\sum_i \bs k_i=0$. 
The integral measures for the phase space are defined as 
 \begin{align}
&[ d x ]=d x_1 d x_2 d x_3\ \delta(1-x_1-x_2-x_3)\ ,
\quad
\notag \\
&[d^2  k ]= \frac{1}{(2\pi)^6}d^2\bs k_{1} d^2\bs k_{2} d^2\bs k_{3}\delta^{(2)}(\bs k_{1}+\bs k_{2}+\bs k_{3})\ .
\label{eq_measure}
\end{align}
The momentum factor $k_L \equiv k^x-ik^y$ is the explicit representation of quark OAM in the proton. 
\subsection{$A$-Form Factor and Helicity-conserved Amplitude at Large $-t$}
\begin{figure}[tpb]

\includegraphics[width=0.7\columnwidth]{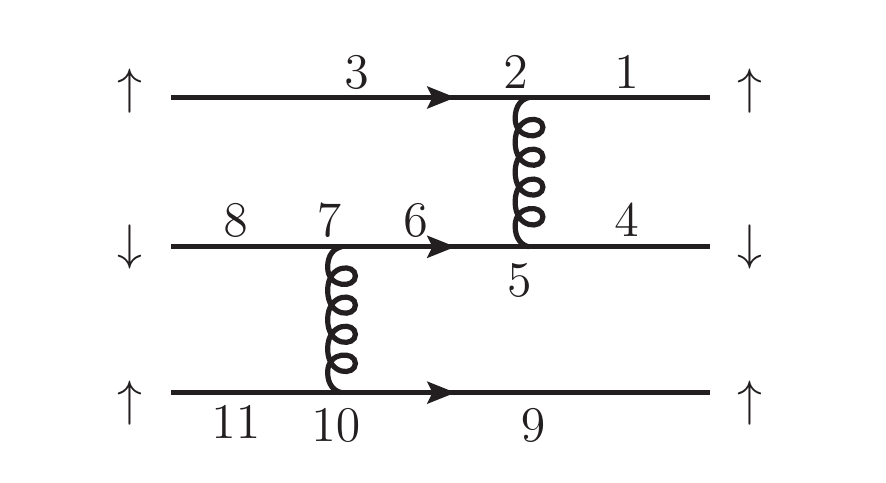}
\includegraphics[width=0.7\columnwidth]{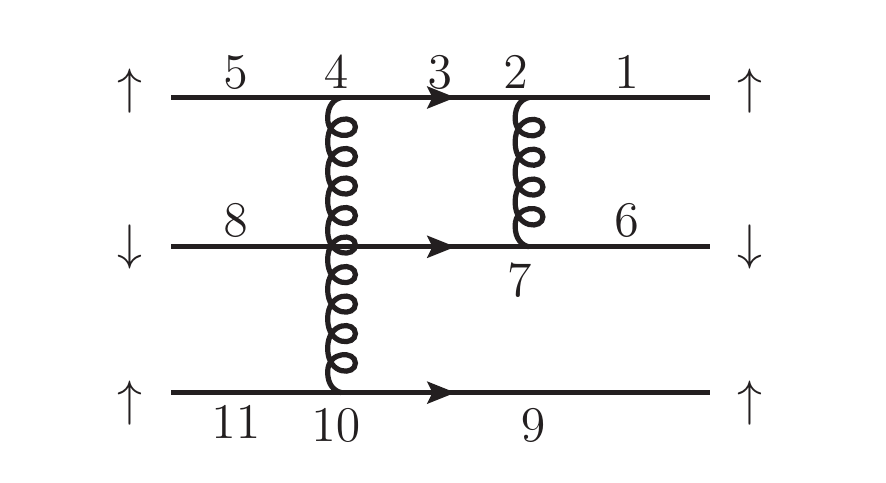}
\includegraphics[width=0.7\columnwidth]{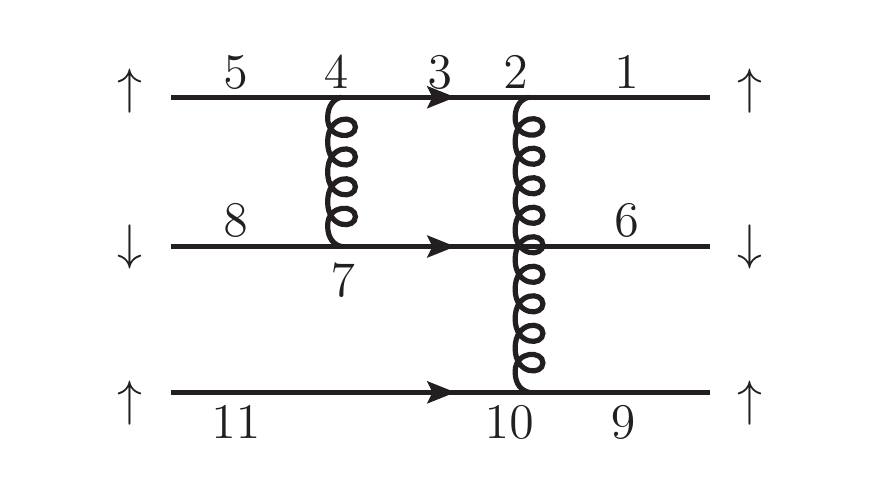}
\caption{Generic diagrams for gravitational form factor calculations. The numbered places represent the insertion of the energy-momentum tensor operator. The three quark lines denote the leading light-cone wave function configuration for the proton state.}
     \label{fig:parton1}
 \end{figure}

As introduced in the Sec.~\ref{sec:helicityamp}, the $A$-GFFs are related to the nucleon helicity-conserved amplitude.
The dominated contributions come from three-quark light-cone  wave function with zero orbital angular momentum:
 \begin{align}
&\langle P'_\uparrow |_{1/2}T^{\mu\nu}_a|P_\uparrow\rangle_{1/2}
 \notag\\
 =&  \int \
   \frac{[d x][d y][d^2 \bs k'][d^2 \bs k]}{24(x_1 x_2 x_3 y_1 y_2 y_3)^{1/2}}
 \psi_1^*
   ( \kappa'_1,\kappa'_2,\kappa'_3)\psi_1
   ( \kappa_1,\kappa_2,\kappa_3) 
   \notag\\
&\times 
 \big\langle  \{y_i P'+\bs k'_i\}\big|_{1/2}
  {T^{\mu\nu}_a}
\big|\{x_i P+\bs k_i\}\big\rangle_{1/2} \ .
 \end{align}
At large momentum transfer, the transverse momenta of the partons in the above amplitude can be neglected. 
Therefore, 
  \begin{align}
&A_a(t)\bar u_\uparrow(P') \gamma^{(\mu} \bar P^{\nu)} 
u_\uparrow(P)
 \notag\\
 =&  \int \
   \frac{[d x][d y]}{96(x_1 x_2 x_3 y_1 y_2 y_3)^{1/2}}
\Phi_3^*(y_1,y_2,y_3)
  \Phi_3(x_1,x_2,x_3)
   \notag\\
&\times \big\langle  uud-udu,\{y_i P'\}\big|
  {T^{\mu\nu}_a}
\big| uud-udu,\{x_i P\}\big\rangle\ ,
\label{eq_twist3proton}
 \end{align}
where $\Phi_3$ is the twist-three light-cone amplitude of proton~\cite{Braun:1999te}:
 \begin{align}
\Phi_3(x_1,x_2,x_3)=2\int[d^2 \bs k]\psi_1(\kappa_1,\kappa_2,\kappa_3)~.
\end{align}
The label $uud$ and $udd$ denote the relevant Fock state configurations in Eq.~(\ref{eq_fock3}):  
\begin{align}
&|uud,\{p_i\} \rangle =\frac{\epsilon_{abc}}{\sqrt{6}}
  \hat u_{a,\uparrow}^\dagger (p_{ 1} ) 
   \hat u_{b,\downarrow}^\dagger (p_{2 })
   \hat d_{c,\uparrow}^\dagger ( p_{3 })\big|0 \big\rangle
   ~,\notag 
   \\
   &
  |udu,\{p_i\} \rangle =\frac{\epsilon_{abc}}{\sqrt{6}}
    \hat  u_{a,\uparrow}^\dagger (p_{ 1} ) \hat d_{b,\downarrow}^\dagger( p_{2 }) \hat u_{c,\uparrow}^\dagger ( p_{3 })\big|0 \big\rangle~.
     \label{eq_partonstate}
\end{align} 
Since the helicity and flavor of  a massless quark are conserved in the fermion line, there is no interaction between these two configurations. Both of them share the same kind of diagrams without any quark line crossing, as presented in Fig.~\ref{fig:parton1}. 
The second configuration in Eq.~(\ref{eq_partonstate}) have additional diagrams with the final $u$ quarks exchanged from Fig.~\ref{fig:parton1}. Therefore, the hard partonic part for the EMT amplitude can be expressed as 
 \begin{align}
H_{a}^{\mu\nu}=2 {\cal H}_{a}^{\mu\nu} +{{\cal H}'}^{\mu\nu}_{a},
\label{eq_hardstru}
\end{align}
where 
\begin{align}
    {\cal H}_a^{\mu\nu}(\{ p_i\})= \big\langle  uud,\{y_i P'\}\big|
  {T^{\mu\nu}_a}
\big| uud,\{x_i P\}\big\rangle\ ,
\label{eq_partonicAmp}
\end{align}
and ${\cal H}'_a$ is obtained from ${\cal H}_a$ by interchanging $y_1$ and $y_3$. 
 \par 
As shown in Fig.~\ref{fig:parton1}, for the quark contribution, the numbered places represent the insertion of the quark EMT operator. Similarly, we can have gluon contribution, where the insertion only appears on the two gluon propagators. The total contribution to the GFF comes from the sum of all these diagrams with all possible insertions. The diagrams in Fig.~\ref{fig:parton1} represent the two different configurations: (1) two gluons attach to the spin-down quark; (2) two gluons attach to the spin-up quark. All other diagrams can be obtained by an exchange of the quark lines $(1\leftrightarrow 3)$. 
 \par

Since the partons are approximately on-shell, we use the covariant perturbation theory in Feynman gauge to compute the partonic amplitudes. 
The technical part of the computation is to evaluate the Dirac structure which involve three fermion lines. In general, they have the following forms:
 \begin{align}
&\bar u_\uparrow(p'_1) \Gamma_1 u_\uparrow(p_1) \ ,\nonumber
\\
&\bar u_\downarrow(p'_2)  \Gamma_2 u_\downarrow(p_2)\ , \nonumber\\
& \bar u_\uparrow(p'_3)  \Gamma_3 u_\uparrow(p_3)\ .
\label{eq_fermion}
\end{align}
Since we only concentrate on the Dirac structure, all irrelevant factors are included in the $\Gamma_{1,2,3}$, which denote the product of $\gamma$-matrix with odd number from three fermion lines presented in the Fig.~\ref{fig:parton1}. All the diagrams share the same color factor,
     \begin{align}
    C_B^2&= \frac{1}{6}\epsilon_{ijk}\epsilon_{i'j'k'} (T^a)_{i'i} (T^a T^b)_{j'j}(T^b)_{k'k}
  =
     \left(\frac{2}{3} \right)^2 \ ,
     \label{eq_color}
     \end{align}
where $T^a$ is the generator in the fundamental representation.
Note that the helicity amplitudes in Eq.~(\ref{eq_fermion}) have a pair of quark lines with zero total helicity. One can apply the following strategy to combine them into a Dirac trace. First, we move out the momentum faction inside the spinor by $u_\lambda(x_i P)=\sqrt{x_i} u_\lambda(P)$, and similarly for $\bar u_\lambda(y_i P')$. The next step is to utilize the identity
\begin{align}
&\bar u_{\uparrow/\downarrow}(P')  \Gamma u_{\uparrow/\downarrow}(P) =\bar u_{\downarrow/\uparrow}(P)  \Gamma_R u_{\downarrow/\uparrow}(P') \ ,
\label{eq_tech1}
\end{align}
where $\Gamma_R$ is obtained from $\Gamma$ by reversing the order of the $\gamma$-matrix. Then we can use the spin density matrix for massless spinor
\begin{align}
u_{\uparrow/\downarrow} (P)\bar u_{\uparrow/\downarrow}(P)=\frac{1}{2} (1\pm \gamma_5) \slashed P\ , 
\label{eq_tech2}
\end{align}
which will turn two fermion lines with opposite helicity of Eq.~(\ref{eq_fermion}) into a Dirac trace. After that, Eq.~(\ref{eq_fermion}) becomes
 \begin{align}
\text{Tr} \bigg[\frac{1}{2}(1+\gamma_5)\slashed P' \Gamma_1 \frac{1}{2}(1+\gamma_5)\slashed P \Gamma_{2R} \bigg]
\ \bar u_\uparrow(P')  \Gamma_3 u_\uparrow(P)\ .
\label{eq_tech3}
\end{align}
Evaluating the trace, contracting the Lorentz indices, and applying the on-shell condition for the spinor $ \slashed P u_\lambda(P)=0,\ \bar u_\lambda(P') \slashed P'=0$, we arrive at the following four Dirac bilinear:
\begin{align}
&E_1
\bar u_\uparrow(P') \bar P^{(\mu} \gamma^{\nu)} u_\uparrow(P) +
E_2
i \bar P^{(\mu} \epsilon^{\nu)\sigma PP'} \bar u_\uparrow(P') \gamma_\sigma u_\uparrow(P)\notag \\ 
+&E_3  \bar u_\uparrow(P')\Delta^{(\mu} \gamma^{\nu)} u_\uparrow(P)
+E_4
i \Delta^{(\mu} \epsilon^{\nu)\sigma PP'} \bar u_\uparrow(P') \gamma_\sigma u_\uparrow(P)\ ,
\end{align}
where $E_i\equiv E_i(\{x_i\},\{y_i\})$. In particular, from the explicit calculation, one can find that $E_3$ and $E_4$ have the form as $x_i-y_i$, and hence the second line with $\Delta$ factor should vanish due to the symmetry between the initial and final state. Furthermore, one can eliminate the Levi-Civita tensor by the identity:
\begin{align}
  &\bar u_\uparrow(P')\gamma^\mu u_\uparrow(P) =  \frac{2i }{-t} \epsilon^{\mu \nu P P' } \bar u_\uparrow(P')\gamma_\nu u_\uparrow(P)\ ,
  \label{eq_tech4}
\end{align} 
which can be verified by using the explicit expression of the Dirac spinors given in Eq.~(\ref{eq_spinor}).
Therefore, one finally obtain the Dirac structure in the proton helicity-conserved amplitude as expected:
\begin{align}
E(\{x_i\},\{y_i\})\bar u_\uparrow(P') \gamma^{(\mu} \bar P^{\nu)} 
u_\uparrow(P)\ .
\end{align}
The hard coefficients for $A_a(t)$ are straightforward to find out.
After summarizing the results, the $A_a(t)$ GFF for the nucleon has the following factorization formula at large momentum transfer:
\begin{align}
A_a( t) =&\int [d x][ dy]\Phi_3^*(y_1,y_2,y_3)
  \Phi_3(x_1,x_2,x_3)\nonumber\\
  &\times \  {\cal A}_a(\{x\} ,\{y\}) \ ,
  \label{A}
 \end{align}
where $\{x\}=(x_1,x_2,x_3 )$, $[d x]= d x_1 d x_2 d x_3\delta(1-x_1-x_2-x_3)$, and $ \Phi_3(x_i)$ is the twist-three light-cone amplitude of the proton~\cite{Braun:1999te}. The hard function can be written as 
 \begin{align}
{\cal A }_a(\{x\} ,\{y\}) =\frac{ 4\pi^2\alpha_s^2  C_B^2}{3t^2  }\left(2{\cal A}_a+{\cal A}'_a\right),
 \end{align}
where ${\cal A}'_a$ is obtained from ${\cal A}_a$ by interchanging $y_1$ and $y_3$.  $C_B=2/3$ is the color factor. For the gluon part, ${\cal A}_g$ can be written as
\begin{align} 
{ \cal  A}_g=&
\frac{x_1+y_1}{\bar x_1\bar y_1 x_1  x_3 y_1  y_3}+
\frac{x_1+y_1}{\bar x_1 \bar y_1 x_1 x_2y_1 y_2 }
+(1\leftrightarrow 3).
\end{align}
For the quark part, the hard coefficient can be expressed as
\begin{align}
{\cal A}_q=&\frac{2}{x_3 y_3 \bar{x}_1^2 \bar{y}_1^2}+\frac{2}{x_2 y_2 \bar{x}_3^2 \bar{y}_3^2}+\frac{2(\bar y_3+\bar{x}_1)-1}{x_1 x_3 y_1 y_3 \bar{x}_1 \bar y_3}
\notag \\&
+\frac{\bar{x}_1-x_1}{x_2 x_3 y_2 y_3 \bar{x}_1 \bar{y}_3} 
+\frac{\bar{y}_1-y_1}{x_2 x_3 y_2 y_3 \bar{x}_3 \bar{y}_1}+(1\leftrightarrow3) \ .
\end{align}
In summary, the quark and gluon $A$-form factors have the same power behavior at large momentum transfer. 

\par

\subsection{Helicity-flip Amplitude at Large $-t$ \label{sec_flip}}
Now we turn to the computation of $B,C$ and $\overline{ C}$ form factors at large $-t$. As shown in Eq.~(\ref{eq_tensor}), they can be extracted respectively from the proton helicity-flip amplitude of EMT with appropriate tensor projections.
The dominant contributions come from the amplitude components with one unit of OAM from either initial or final state nucleons. Since they yield the same contribution, we only focus on one of them:
\begin{align}
&\langle P'_\uparrow |^{l'_z=0}_{1/2}T^{\mu\nu}_a|P_\downarrow\rangle_{1/2}^{l_z=-1}
 \notag\\
 =&  \int \
   \frac{[d x][d y][d^2 \bs k'][d^2 \bs k]}{24(x_1 x_2 x_3 y_1 y_2 y_3)^{1/2}}
  \big[  k_{1L} \psi_3+  k_{2L} \psi_4 \big]
   ( \kappa_2,\kappa_1,\kappa_3) 
   \notag\\
&\times \psi_1^*
   ( \kappa'_1,\kappa'_2,\kappa'_3)
 \big\langle  \{y_i P'+\bs k'_i\}\big|_{1/2}
  {T^{\mu\nu}_a}
\big|\{x_i P+\bs k_i\}\big\rangle_{1/2} \ ,
 \end{align}
where the factors $k_{L}=k^x-i k^y$ are manifestations of the quark OAM with $l_z=-1$ from the initial state proton. The partonic amplitudes have the same structure as that in Eq.~(\ref{eq_hardstru}) except that the transverse momenta of the initial state partons can not be ignored.
Otherwise, the amplitudes will vanish due to the OAM factor $k_{L}$ in the phase-space integral. Instead, we need to evaluate the following partonic amplitude:
\begin{align}
\big\langle  uud,\{y_i P'\}\big|\Gamma_{\mu\nu}
  T^{\mu\nu}_a
\big| uud,\{x_i P+\bs k_i\}\big\rangle,
\end{align} 
where we have used the label $\Gamma_{\mu\nu}$ to represent the tensor projection for the $B,C,$ or $\overline C$ form factors in Eq.~(\ref{eq_tensor}). The partonic state $|uud\rangle$ has been defined in Eq.~(\ref{eq_partonstate}). Similar to the $A$-form factor, there are three fermion lines responsible for the three-valence-quark configuration for each diagram in Fig.~\ref{fig:parton1}. These perturbative diagrams share the same color factor as that in Eq.~(\ref{eq_color}). They also contain the similar Dirac structure as that in Eq.~(\ref{eq_fermion}) so that we apply the similar strategy for the Dirac algebra, following Eqs.~(\ref{eq_tech1}-\ref{eq_tech3}) and Eq.~(\ref{eq_tech4}). 
In order to obtain the leading contribution at large $-t$, we should expand the transverse momenta in terms of $\bs k_i/\sqrt{-t}$. The linear terms of $\bs k_i$ will contribute at this order. In particular, one can apply the following formula for the spinor:
\begin{align}
u_\lambda(x_i P+\bs k_i)&\approx u_\lambda(x_i P)+\frac{\slashed k_i\slashed P'}{x_i P'  \cdot P }  u_\lambda(x_i P) \ ,
\end{align} 
which leads to a linear expansion term. 

Applying the transverse momentum expansion of the partonic scattering, we obtain the following expression for the linear expansion term,  
\begin{align}
&{\cal H}_1(\{x_i\},\{y_i\})
\bar u_\uparrow(P') \slashed {\bs k}_{1}  u_\uparrow(P)
\notag \\
&+{\cal H}_3(\{x_i\},\{y_i\})\bar u_\uparrow(P') \slashed {\bs k}_{3}  u_\uparrow(P)\ ,
\end{align}
where $\bs k_{2 }=-\bs k_{1}- \bs k_{3}$ has been used to simplify the final results. To obtain the helicity-flip like structure, one can use the identity 
$
\gamma^i u_\uparrow(p)=(\delta^{ix}+ i \delta^{iy}) u_\downarrow(p) 
$ for $i=x,y$,
and hence the bilinear spinor can be further reduced as
$\bar  u_\uparrow(P')\slashed {\bs k}_{i}  u_\uparrow(P) =k_{iR} \bar  u_\uparrow(P') u_\downarrow(P)$.
Therefore, the GFFs can be expressed as
\begin{align}
&\langle P'_\uparrow |^{l'_z=0}_{1/2} \Gamma_{\mu\nu}T^{\mu\nu}_a|P_\downarrow\rangle_{1/2}^{l_z=-1}
 \notag\\
 =& \bar  u_\uparrow(P') u_\downarrow(P) \int [d x][d y][d^2 \bs k'][d^2 \bs k]\psi_1^*
   ( \kappa'_1,\kappa'_2,\kappa'_3)
   \notag\\
&\times \big[  k_{2L} \psi_3 ( \kappa_2,\kappa_1,\kappa_3)+  k_{1L} \psi_4 ( \kappa_2,\kappa_1,\kappa_3) \big]
\notag \\
&\times [ k_{1R}{ H}_1(\{x_i\},\{y_i\})+k_{3R} { H}_3(\{x_i\},\{y_i\}) ]
 \end{align}
where $k_{R/L}=k^x\pm i k^y$. The hard parts have the form as $H_i=2{\cal H}_i+{\cal H}'_i$ and ${\cal H}'_i={\cal H}_i(y_1\leftrightarrow y_3)$ where the $x_i,y_i$-dependences from the phase integral are also included. On the other hand, from the property of the transverse-momentum integral, one has $
\int [d^2 \bs k]  k_{iL}  k_{jR}  f(\bs k_i\cdot\bs  k_j)
=\int  [d^2 \bs k] \  \bs k_i\cdot \bs k_j f(\bs k_i\cdot\bs  k_j)
$.  The transverse-momentum dependence from the initial state can be absorbed into the twist-4 light-cone distribution amplitudes of nucleon by the following relations~\cite{Belitsky:2002kj}: 
\begin{align}
\Phi_4 (  x_1 ,x_2,x_3)= & 2\int \frac{ [d^2 \bs k]}{M x_3}\  \bs k_{3}\cdot \big [ \bs  k_{1}  \psi_3 (\kappa_2,\kappa_1,\kappa_3)
\notag \\& \qquad +\bs  k_{2}  \psi_4 (\kappa_2,\kappa_1,\kappa_3)\big]\ ,
\notag \\
\Psi_4 ( x_2,x_1 ,x_3)=&2\int \frac{ [d^2 \bs k]}{M x_1}\  \bs k_{1}\cdot \big [ \bs  k_{2} \psi_3 (\kappa_2,\kappa_1,\kappa_3)
\notag \\&\qquad 
+\bs  k_{1}  \psi_4(\kappa_2,\kappa_1,\kappa_3)  \big] \ .
\end{align}
The normalizations of these twist-4 amplitudes are the same as those in~\cite{Braun:2000kw}. 
Similarly, the transverse momenta of the final wave function can also be integrated out as in Eq.~(\ref{eq_twist3proton}) and thus the twist-3 distribution amplitude of the final nucleon is obtained. Therefore, the GFFs at large momentum transfer have the following general factorization structure:
\begin{align}
&\langle P'_\uparrow |^{l'_z=0}_{1/2} \Gamma_{\mu\nu}T^{\mu\nu}_a|P_\downarrow\rangle_{1/2}^{l_z=-1} 
\notag \\&=  \int [d x][d y] \left\{ x_3 \Phi_4(x_1,x_2,x_3)  H_{\Phi a}(\{x\},\{y\})\right.\nonumber\\
&
\left.+x_1\Psi_4(x_2,x_1,x_3)  H_{\Psi a}(\{x\},\{y\})
\right\}\  \Phi_3(y_1,y_2,y_3) \ ,
\label{eq_twist4structure}
\end{align} 
where $\Phi_3$ is the twist-3 distribution amplitude and  $\Phi_4,\Psi_4$ are the twist-4 distribution amplitudes defined above. They embrace the non-perturbative effects of the GFFs at large $-t$. $H_{\Phi a}, H_{\Psi a}$ are the hard coefficients and can be calculated order by order in principle. 
\par Although we derived the above formula at the lowest order perturbation theory, we expect that it still takes the same form when high-order $\alpha_s$-corrections are considered. We notice that the bare quark or gluon GFFs suffer from UV-divergence and need renormalization at higher orders~\cite{Hatta:2018sqd,Tanaka:2018nae}. The operator mixing is needed and the trace anomaly $F^2$ plays an important role in the renormalization of EMT operator~\cite{Hatta:2018sqd,Tanaka:2018nae,Freedman:1974gs,Nielsen:1975ph,Nielsen:1977sy,Collins:1976yq,Tarrach:1981bi,Ji:1995sv}, which is not present in our leading order calculation. As shown in Sec.~\ref{sec:SFF}, the $\langle P'|F^2| P\rangle$ for nucleon at large $-t$ is also related to the helicity-flip amplitude. Hence, the factorization structure of $B,C,\overline { C}$ form factors in Eq.~(\ref{eq_twist4structure}) should not be altered by the operator mixing. 

\subsection{$\overline C$ Form Factors and Check on the Cancellation between Quarks and Gluons\label{sec:cbarcheck}}
In the last section, we have explained the general procedure on the  calculation of the $B$, $C$ and $\overline{C}$ form factors from the helicity-flip amplitude with different tensor projections. In particular, $\overline{C}$ form factor is actually determined by the traceless feature $g_{\mu\nu}T_{q,g}^{\mu\nu}=0$ at this order. It is an important cross check that the total contribution of $\overline{C}$ from the quarks and gluons vanishes because of the current conservation.  In order to show that from the explicit calculation, we define the following partonic amplitude for $\overline{C}$-GFFs:
\begin{align}
\overline{ {\cal C}}_a\equiv\frac{\big\langle  uud,\{y_i P'\big|\Gamma_2^{\mu\nu}
  T^{\mu\nu}_a
\big| uud,\{x_i P+\bs k_i\}\big\rangle}{48 (x_1 x_2 x_3 y_1 y_2 y_3)^{1/2}\Delta^2\bar u_\uparrow(P')u_\downarrow(P)} \ .
\end{align} 
where $\Gamma_2^{\mu\nu}=\Delta^\mu\Delta^\nu-\Delta^2 g^{\mu\nu}/4$ has been defined in Eq.~(\ref{eq_tensor}).
\par 
Following the strategy presented in the last section, we can go through the detailed but straightforward computation on the perturbative diagrams in Fig. \ref{fig:parton1} and check explicitly the cancellation of $\overline{C}$ form factor contributions from the quarks and gluons. In the followings, we will list the contributions of $\overline {\cal C}_{q,g}$ and an overall factor is implied for brevity:
\begin{align}
-\frac{2\pi^2  \alpha_s^2 C_B^2}{3(-t)^2}\ .
\end{align} 
First, we summarize the contribution to the gluon GFF, 
\begin{eqnarray}
\overline {\cal C}_{g1}^{(1)}&=&\frac{(x_3 k_{1R} +\bar x_1 k_{3R} )(x_1\bar x_1+y_1\bar y_1)}{\bar x_1^2\bar y_1x_1x_3^2y_1y_3^2}-\frac{(\bar x_1+y_1) k_{3R}}{\bar x_1^2\bar y_1x_1x_3y_1y_3}\ ,\nonumber\\
\overline {\cal C}_{g2}^{(1)}&=&\frac{((x_3\bar x_3 + y_3\bar y_3) k_{1R} }{\bar x_3\bar y_3^2x_1^2x_3y_1y_3} -\frac{(\bar y_3 - x_3) k_{3R}}{\bar x_3\bar y_3^2x_1x_3y_1y_3}\ ,\nonumber\\
\overline {\cal C}_{g1}^{(2)}&=&-\frac{(x_2\bar x_2+y_2\bar y_2)(x_3 k_{1R}-x_1 k_{3R})}{x_1x_2x_3^2y_2y_3^2\bar x_2\bar y_2} \ ,\nonumber\\
\overline {\cal C}_{g2}^{(2)}&=&\frac{(x_3\bar x_3 + y_3\bar y_3) k_{1R} }{x_1x_2x_3y_2y_3\bar x_3\bar y_3^2}- \frac{(\bar x_3-y_3 ) k_{3R}}{x_2x_3y_2y_3\bar x_3\bar y_3^2}\ , \nonumber\\
\overline {\cal C}_g^{(3)}&=&- \frac{(y_3 + \bar x_3)  k_{3R}}{x_2x_3y_2y_3\bar x_3^2\bar y_3} \ .
\end{eqnarray}
The total contribution will be obtained by adding the above together and apply $(1\leftrightarrow 3)$.

To compute the quark contributions from the diagrams of Fig.~\ref{fig:parton1}, we notice that the total contribution from $g^{\mu\nu}$ term in $\Gamma_2^{\mu\nu}$ cancel out among the diagrams. Therefore, we only need to take into account the $\Delta^\mu\Delta^\nu$ term in $\Gamma_2^{\mu\nu}$ in the projection of these diagrams. For the first diagram, we have
\begin{eqnarray}
\overline {\cal C}_{q1}^{(1)}&=&\frac{-(x_3 k_{1R} +\bar x_1 k_{3R})}{\bar x_1^2x_3^2y_3^2}\ ,~~\overline {\cal C}_{q2}^{(1)}=\frac{ k_{1R}}{\bar x_1^2\bar y_1x_1x_3y_3}\ ,\nonumber\\
\overline {\cal C}_{q3}^{(1)}&=&\frac{-(x_3 k_{1R} +\bar x_1 k_{3R})}{\bar x_1\bar y_1x_3^2y_3^2}\ , ~~\overline {\cal C}_{q4}^{(1)}=\frac{ {-\bar y_2} (x_3 k_{1R} +\bar x_1 k_{3R})}{\bar x_1x_1x_3^2y_1y_3^2}\ ,\nonumber\\
\overline {\cal C}_{q5}^{(1)}&=&\frac{ k_{1R}}{\bar x_1 x_1^2x_3y_1y_3}\ , ~~\overline {\cal C}_{q7}^{(1)}=\frac{- k_{3R}}{\bar y_3x_1x_3^2y_1y_3}\ ,\nonumber\\
\overline {\cal C}_{q6}^{(1)}&=&\frac{ (y_3-x_1)(x_1(x_3 k_{1R} +\bar x_1 k_{3R} )-x_3y_3 k_{1R})}{\bar x_1 \bar y_3x_1^2x_3^2y_1y_3^2}\ ,\nonumber\\
\overline {\cal C}_{q8}^{(1)}&=&\frac{\bar x_2(x_1 k_{3R}-x_3 k_{1R} )}{\bar y_3 x_1^2x_3^2y_1y_3}\ ,~~\overline {\cal C}_{q9}^{(1)}=\frac{(x_1 k_{3R}-x_3 k_{1R})}{\bar x_3\bar y_3 x_1^2x_3y_1} \ , \nonumber\\
\overline {\cal C}_{q10}^{(1)}&=&\frac{- k_{3R}}{\bar x_3\bar y_3 ^2x_1x_3y_1}\ , ~~\overline {\cal C}_{q11}^{(1)}= \frac{(x_1 k_{3R}-x_3 k_{1R})}{\bar y_3^2x_1^2x_3y_1}\ .
\end{eqnarray}
It is interesting to find out that the total contribution from the quark and gluons takes the following expression,
\begin{eqnarray}
\overline {\cal C}_{q,g}^{(1)}&=&\frac{2(\bar x_1-y_1)k_{1R}}{\bar x_1\bar y_1x_1^2x_3y_1y_3}- \frac{2(\bar x_3-y_3)k_{3R}}{ \bar x_3\bar y_3x_1x_3^2y_1y_3}\ ,
\end{eqnarray}
which is anti-symmetric under the exchange of $(1\leftrightarrow 3)$. Therefore, the cancellation of $\overline{C}$ form factor between the quark and gluon contributions happens when we adding the first diagram of Fig.~\ref{fig:parton1} and its mirror. 

For the second diagram in Fig.~\ref{fig:parton1}, we have
\begin{eqnarray}
\overline {\cal C}_{q1}^{(2)}&=&\frac{\bar y_1 (x_3k_{1R}-x_1k_{3R})}{x_1x_2x_3^2y_2y_3^2}\ ,~~\overline {\cal C}_{q4}^{(2)}=\frac{- k_{3R}}{\bar y_3 x_2x_3^2y_2y_3} \ ,\nonumber\\
\overline {\cal C}_{q3}^{(2)}&=&\frac{(y_3-x_2 )(x_1k_{3R}-x_3k_{1R})}{\bar y_3x_1x_2x_3^2y_2y_3^2}\ ,~~\overline {\cal C}_{q2}^{(2)}=0 \ ,\nonumber\\
\overline {\cal C}_{q5}^{(2)}&=&\frac{\bar x_1(x_1k_{3R}-x_3k_{1R})}{\bar y_3x_1x_2
x_3^2y_2y_3} \ ,~~\overline {\cal C}_{q6}^{(2)}=\frac{( x_3k_{1R}-x_1k_{3R} )}{\bar x_2x_1x_3^2y_3^2}\ , \nonumber\\
\overline {\cal C}_{q7}^{(2)}&=&0 \ ,~~\overline {\cal C}_{q8}^{(2)}=\frac{( x_3k_{1R}-x_1k_{3R})}{\bar y_2x_1
x_3^2y_3^2}\ ,\nonumber\\
\overline {\cal C}_{q9}^{(2)}&=&\frac{(x_1k_{3R}-x_3k_{1R} )}{\bar x_3\bar y_3x_1x_2x_3
y_2} \ ,~~\overline {\cal C}_{q10}^{(2)}=\frac{-x_1k_{3R}}{\bar x_3 \bar y_3^2x_2x_3y_2} \ ,\nonumber\\
\overline {\cal C}_{q11}^{(2)}&=&\frac{(x_1k_{3R}-x_3k_{1R} )}{\bar y_3^2x_1x_2x_3y_2}\ . 
\end{eqnarray}
We can also sum up quark and gluon contributions in the second diagram,
\begin{equation}
    \overline{\cal C}_{q,g}^{(2)}=\frac{( y_3-\bar x_3 )k_{3R}}{\bar x_3\bar y_3x_2x_3^2y_2y_3} \ ,
\end{equation}
which does not vanish, even considering $(1\leftrightarrow 3)$ symmetry. However, we will show this is canceled by the contributions from the third diagram.

The quark contributions from the third diagram of Fig.~\ref{fig:parton1} are very simple, 
\begin{eqnarray}
\overline {\cal C}_{q2}^{(3)}&=&\frac{k_{3R}}{\bar x_3 x_2x_3^2y_2y_3} \ , ~~\overline {\cal C}_{q10}^{(3)}=\frac{k_{3R}}{\bar x_3^2\bar y_3x_2x_3y_2} \ ,
\end{eqnarray}
and all others vanish. 
Adding them together, we have,
\begin{equation}
    \overline{\cal C}_{q,g}^{(3)}=\frac{(\bar x_3 - y_3)k_{3R} }{\bar x_3\bar y_3x_2x_3^2y_2y_3} \ .
\end{equation}
This exactly cancels that from $\overline{\cal C}_{q,g}^{(2)}$. 

To summarize, we have shown that the total $\overline{C}$ form factor factor vanishes when adding all contributions from the quarks and gluons. This provides an important cross check of our derivations. Moreover, from above results, we can obtain the factorization formula of $\overline C_a$ form factor for quark or gluon respectively. It takes the same form as shown in the Eq.(\ref{eq_twist4structure}):
\begin{align}
&\overline C_a(t) =  \int [d x][d y] \left\{ x_3 \Phi_4(x_1,x_2,x_3)  \overline { C}_{\Phi, a}(\{x\},\{y\})\right.\nonumber\\
&
\left.+x_1\Psi_4(x_2,x_1,x_3)\overline  { C}_{\Psi, a}(\{x\},\{y\})
\right\}\  \Phi_3(y_1,y_2,y_3) \ ,
\label{eq_cbarfac}
\end{align} 
where  the hard functions have the following structure:
 \begin{align}
&\overline{ C}_{\Phi, a}=2 \overline{\cal C}_{\Phi, a} +\overline{\cal C}'_{\Phi,  a},
\notag \\
&\overline{ C}_{\Psi, a}=\overline{ C}_{\Phi, a}(1\leftrightarrow 3),
\label{eq_cbarfac2}
\end{align}
and $\overline {\cal C}'$ is obtained from $\overline {\cal C}$ by interchanging $y_1$ and $y_3$.  
At the leading order perturbation theory, we obtain
\begin{align}
\overline {\cal C}_{\Psi, q}=&-\overline {\cal C}_{\Psi, g }
\notag \\
=&\frac{2\pi^2 M C_B^2\alpha_s^2}{3(-t)^2}\Big [x_3  (x_1\bar x_1+y_1\bar y_1) K_1
\notag \\&
-x_1 y_1 (\bar x_3+y_3) \tilde K_1
\notag \\
&
+(x_3\bar x_3 + y_3\bar y_3)(\bar x_3 \tilde K_1 +\tilde K_4+\tilde K_5)
\notag\\
&+(x_2\bar x_2+y_2\bar y_2)x_3  (\tilde K_2-K_2) 
\notag \\ 
&-[\bar x_1 (\bar x_1-y_1 ) 
+\bar y_1(y_1+ \bar x_1)]K_3/\bar y_1 
\Big ] \ .
\label{eq_cbarfac3}
\end{align}
The functions $K_{i}$ are defined as
\begin{align}
&K_1=\frac{1}{x_1 x_3^2 y_1 y_3^2 \bar{x}_1^2 \bar{y}_1},
\quad
K_2=\frac{1}{x_1 x_2 x_3^2 y_2 y_3^2 \bar{x}_2 \bar{y}_2},\nonumber\\
&K_3=\frac{1}{x_1 x_2 y_1 y_2 \bar{x}_1^2 \bar{y}_1}
,
\quad
K_4=\frac{1}{x_1 x_3^2 y_1 y_3 \bar{x}_1 \bar{y}_1^2},
\notag \\
&K_5=\frac{1}
{x_1 x_2 x_3 y_1 y_2 \bar{x}_1 \bar{y}_1^2}, \quad  \tilde K_i=K_i(1\leftrightarrow3)\ .
\label{eq_Kfunction}
\end{align}

\subsection{$B$ and $C$ Form Factors at Large $-t$}
Similar to the $\overline C$-form factor, the $B$ and $C$-form factors for quarks and gluons come from the helicity-flip amplitude and can be calculated by the same analysis explained in Sec.~\ref{sec_flip} at large momentum transfer. They all follow the same factorization structure as in the Eq.~(\ref{eq_twist4structure}) or Eq.~(\ref{eq_cbarfac}). In this subsection, we will summarize these results.

\paragraph{Gluon form factor}
With the above analysis, we carry out a detailed derivation for all the diagrams of Fig.~\ref{fig:parton1} and $C_g(t)$ can be factorized into,
\begin{align}
&C_g(t) =  \int [d x][d y] \left\{ x_3 \Phi_4(x_1,x_2,x_3)  { C}_{\Phi,g}(\{x\},\{y\})\right.\nonumber\\
&
\left.+x_1\Psi_4(x_2,x_1,x_3) { C}_{\Psi, g}(\{x\},\{y\})
\right\}\  \Phi_3(y_1,y_2,y_3) \ .
\end{align} 
The hard functions can be written as
\begin{align}
{ C}_{\Psi,g}&=2 {\cal C}_{\Psi,g} +{\cal C}'_{\Psi,g}~, 
\notag \\
{ C}_{\Phi,g}&={ C}_{\Psi,g}(1\leftrightarrow3)~,
\end{align}
where ${\cal C}'$ is obtained from ${\cal C}$ by interchanging $y_1$ and $y_3$.  From the detailed calculations of the diagrams in Fig.~\ref{fig:parton1}, we obtain
\begin{align}
{\cal C}_{\Psi,g }(\{x\},\{y\})=\frac{ C_B^2  M^2 }{24(-t)^3} (4\pi \alpha_s)^2{\cal H}_{g}\bigg \vert_{\text{upper sign}}, 
\label{eq:Cghard}
\end{align} 
where
 \begin{align}
&{\cal H}_{g}(\{x\},\{y\})= %
  \nonumber\\
&~ \bigg[
x_3 K_1 
 \left( x_1
   \bar{x}_1+y_1 y_2 \mp 2 y_3 \bar{x}_1\right)
+
\bar x_3 \tilde K_1 
 \left(x_3 \bar{x}_3+y_3 \bar{y}_3\right)  
\notag \\& ~
+
x_3 (\tilde K_2- K_2) 
\left (x_2 \bar{x}_2 + y_2 \bar{y}_2
\right)
+ 
K_3
 \bigl(\mp 2 \bar{x}_1-y_1\bigr )
\notag \\& ~
+x_3 (K_4 + K_5)
\bigl (x_1\mp2 \bar{y}_1\bigr)
+
(\tilde K_4+\tilde K_5)
 ( x_3 \bar{x}_3+ y_3 \bar{y}_3  )
 \bigg],
 \label{eq:ghard}
 \end{align}
and functions $K_{i}$ defined in Eq.(\ref{eq_Kfunction}).

\par 
Likewise, the gluonic $B$-form factor has the similar factorization formula:
\begin{align}
&B_g(t) =  \int [d x][d y] \left\{ x_3 \Phi_4(x_1,x_2,x_3)  { B}_{\Phi,g}(\{x\},\{y\})\right.\nonumber\\
&
\left.+x_1\Psi_4(x_2,x_1,x_3) { B}_{\Psi, g}(\{x\},\{y\})
\right\}\  \Phi_3(y_1,y_2,y_3) \ ,
\end{align} 
 where the hard function ${\cal H}$ for $B_g$ is 
 \begin{align}
{ B}_{\Psi,g}&=2 {\cal B}_{\Psi,g} +{\cal B}'_{\Psi,g}~, 
\notag \\
{ B}_{\Phi,g}&={ B}_{\Psi,g}(1\leftrightarrow3)~,
\end{align}
and ${\cal B}'$ is obtained from ${\cal B}$ by interchanging $y_1$ and $y_3$.  Detailed calculation gives
\begin{align}
{\cal B}_{\Psi,g }(\{x\},\{y\})=-\frac{ C_B^2  M^2 }{6(-t)^3} (4\pi \alpha_s)^2 {\cal H}_{g}\bigg \vert_{\text{lower sign}} \ ,
\label{eq:Bghard}
\end{align}
at the leading order.

\paragraph{Quark form factor} We turn to the case of the quark form factors. Similar analysis and derivation show that $C_q(t)$ can be factorized as
\begin{align}
&C_q(t) =  \int [d x][d y] \left\{ x_3 \Phi_4(x_1,x_2,x_3)  { C}_{\Phi,q}(\{x\},\{y\})\right.\nonumber\\
&
\left.+x_1\Psi_4(x_2,x_1,x_3) { C}_{\Psi, q}(\{x\},\{y\})
\right\}\  \Phi_3(y_1,y_2,y_3) \ ,
\end{align} 
where $\Psi_4$ and $\Phi_4$ are the twist-four distribution amplitude of the proton \cite{Braun:2000kw}. The hard fucntions have the structure:
\begin{align}
{ C}_{\Psi,q}&=2 {\cal C}_{\Psi,q} +{\cal C}'_{\Psi,q}~, 
\notag \\
{ C}_{\Phi,q}&={ C}_{\Psi,q}(1\leftrightarrow3)~,
\end{align}
where ${\cal C}'$ is obtained from ${\cal C}$ by interchanging $y_1$ and $y_3$.  From the detailed computation of diagrams in Fig.~\ref{fig:parton1}, we obtain
  \begin{align}
{\cal C}_{\Psi,q }(\{x\},\{y\})=\frac{ C_B^2  M^2 }{48(-t)^3} (4\pi \alpha_s)^2  {\cal H}_{qC}(\{x\},\{y\})~.
 \end{align}
 \par
Similarly, the $B$-form factor for quarks follow the same form of factorization formula at large $-t$ as the gluonic one and the corresponding hard fucntions have the same structure.  Explicit calculation yields
  \begin{align}
&{\cal B}_{\Psi,q}(\{x\},\{y\})=-\frac{ C_B^2  M^2 }{12(-t)^3} (4\pi \alpha_s)^2  {\cal H}_{qB}(\{x\},\{y\})~.
 \end{align}
\par 
The functions ${\cal H}_{qC}$ and ${\cal H}_{qB}$ can be unifiedly described  by the function ${\cal H}_{q}$ with different signs chosen. The upper sign is for the $C_q$, while the lower sign is for the $B_q$, i.e. 
\begin{align}
{\cal H}_{qC}={\cal H}_{q} \bigg \vert_{\text{upper sign}},\quad
{\cal H}_{qB}={\cal H}_{q} \bigg \vert_{\text{lower sign}},
\end{align}
where
\begin{align}
{\cal H}_{q}=&\left(\bar{y}_1-y_1\pm2\right)
   x_3 (G_5-G_1+\tilde G_4)
    \notag \\
   &
   + \left(\bar{y}_2-y_2\pm2\right) [x_3(G_6- \tilde G_6 - G_2) -\bar{x}_3 \tilde G_2]
   \notag \\
   &-\left(\bar{y}_3-y_3\pm2\right)  [\bar{x}_3 \tilde  G_1
   +x_3 (\tilde  G_5 + G_4)]
    \notag \\
   &+\left(\bar{x}_2-x_2\pm 2 \right) x_3(G_6-\tilde G_6+\tilde G_3 -  G_3)
   \nonumber \\ 
    &-\left(\bar{x}_1-x_1\pm2\right) x_3  G_1 +2 
  (G_1+G_2) x_3 y_3/x_1 
    \notag \\ &
-  \left(\bar{x}_3-x_3\pm 2\right) [x_3 (G_4-y_1 \tilde  G_5/ \bar y_1) +\bar x_3 \tilde G_1 ]
   \nonumber \\ 
    &
+  
   [  (\bar{x}_1-x_1\pm 2) (\pm 2-x_1 ) -3
   ]  [\tilde  G_4 - y_3 G_5/\bar y_3] x_3/\bar x_1 
  \notag \\ &
  +2  (  y_1x_3/\bar x_1\tilde G_5+ G_7)  \mp4  x_3(\tilde G_4 +\tilde  G_3+y_1\tilde  G_5/\bar y_1) 
   \notag\\
 &+   [ \pm 
   2\left(\bar{y}_3-y_3\right)-(\bar{x}_2-x_2)]x_3 G_5/\bar y_3
    \notag \\ &
-[ \pm 
   2\left(\bar{y}_1-y_1\right)-(\bar{x}_2-x_2)]x_3 \tilde G_5/\bar y_1
   \notag\\
&    + 
   \left[(\bar{x}_3-x_3)\pm
   2\left(y_1-\bar{y}_1\right)\right]\bar{x}_3 \tilde G_2/ \bar y_1 
  \notag \\ 
 &+  [\left(x_1-\bar{x}_1\right) \left(\pm 2
   y_3-x_1\right)
    \notag \\ 
 &+\left(y_3-\bar{y}_3\right) \left(\pm 2 x_1-y_3\right)] x_3 G_2/(x_1 \bar y_3).
\end{align}
The functions $G_{i}$ are defined as
 \begin{align}
&G_1=\frac{1}{x_3^2 y_3^2 \bar{x}_1^2 \bar{y}_1},\quad G_2=\frac{1}{ {x_1 x_3^2 y_1 y_3^2 \bar{x}_1}},
\notag\\
&
G_3=\frac{1}{x_1^2 x_3^2 y_1 y_3 \bar{y}_3},
\quad
G_4=\frac{x_1 y_1+x_2 y_2}{x_1^2 x_2 x_3
   y_1 y_2 \bar{x}_3 \bar{y}_3^2},
   \notag\\
&G_5=\frac{1}{x_1 x_2 x_3^2 y_3^2
  y_2}
  ,\quad
 G_6= \frac{1}{x_1
   x_3^2 y_3^2 \bar{x}_2 \bar{y}_2}
 ,\quad
 \notag\\
&
 G_7=\frac{1}{\bar x_1^2\bar y_1x_2x_1y_2}  \ ,\quad  \tilde G_i=G_i(1\leftrightarrow3)\ .
\end{align}
\par 
There are several points we would like to comment on the above results. First, one can carry out the consistent check by calculating the $\overline C_a$ from factor
\begin{align}
\overline{C}_a(t)=& \frac{-t}{16 M^2 } \big [B_a(t)-12 C_a(t) \big]\ .
\label{eq:GFFrelation}
\end{align}
We have checked that the obtained results are the same as those in Eq.~(\ref{eq_cbarfac},\ref{eq_cbarfac2},\ref{eq_cbarfac3}) and hence the total $\overline C$ GFF vanishes as expected, see also, the relation between the total $B$ and $C$ form factors of Eq.~(\ref{eq:bcrelation}).

Second, the above results confirm the power counting for the GFFs: $C_{q,g}, B_{q,g}\sim 1/(-t)^3$, $\overline C_{q,g}\sim 1/(-t)^2$. Compared with the $A$-form factor, the $B,C$-form factor are $1/(-t)$-suppressed due to the quark OAM for the helicity flip amplitude. $\overline C_{q,g}$ also come from the helicity-flip amplitude, but with one power of $(-t)$ enhanced compared to $B_{q,g},C_{q,g}$. This is because the parameterizations of the GFFs.

Third, it is interesting to see that the perturbative coefficients for $B_{q,g},C_{q,g}$ form factor can be described by one function ${\cal H}_{q,g}$ for quark and gluon, respectively:  $B_{q,g}$ and $C_{q,g}$ can be related to each other only by reversing the sign of some terms. In addition, the hard coefficients of $B_{q,g},C_{q,g},\overline C_{q,g}$ GFFs suffer from the end-point singularity when $x_i,y_i \rightarrow 0$. 
This effect comes from the soft partons carrying small fraction of the longitudinal momentum.  Since the longitudinal and transverse momenta of the partons are comparable, we are not able to perform the intrinsic transverse momentum expansion in the end point region. One should develop a rigorous framework to factorize and resum these soft parton contributions in the GFFs, which however is beyond the scope of the current paper. Nonetheless, we would like to point out some observations on the end point behaviors from our results. From Eqs.~(\ref{eq:Cghard},\ref{eq:ghard},\ref{eq:Bghard}), we find that the end point singularities of the gluonic $B$ and $C$ GFFs can be related by
\begin{align}
B_g \bigg \vert_{\text{singular}}=-4C_g\bigg \vert_{\text{singular}},
\end{align}
and hence
\begin{align}
\overline{ C}_g\bigg \vert_{\text{singular}}=& \frac{-t}{16 M^2 }4B_g\bigg \vert_{\text{singular}}
\end{align} 
from Eq.~(\ref{eq:GFFrelation}).
This relation dose not hold for the quark contributions and the end point singularity does not cancel in the total quark and gluon contributions. However, from $\overline C_q=-\overline C_g$, we also have
\begin{align}
\overline{ C}_q\bigg \vert_{\text{singular}}=- \frac{-t}{16 M^2 }4B_g\bigg \vert_{\text{singular}}&,
\notag \\ 
(B_q-12 C_q)\bigg \vert_{\text{singular}}=4B_g\bigg \vert_{\text{singular}}&.
\end{align} 

One more point, the GFFs can be obtained from the GPDs $H_a,E_a$ by taking the first moment. For the gluonic GFFs, we have carried out a consistent check by using  the gluon GPDs at large $-t$ presented in \cite{Sun:2021pyw}. For the $A_q$ GFF, we use the quark GPDs $H_q$ in \cite{Hoodbhoy:2003uu} to check the result, whereas for the other quark GFFs, we compute the quark GPD $E_q$ by using the methods introduced in the previous sections. We list them in the Appendix for reference. The results are all consistent.

\section{Numerical results\label{sec:numerical}}
In this section, we will present numeric results for the gluon and quark GFFs of the nucleon at large $(-t)$. To achieve that, the non-perturbative inputs of the twist-three and twist-four nucleon light-cone distribution amplitudes (DAs) are needed~\cite{Braun:1999te,Braun:2000kw}. There have been progresses in the determination of the nucleon DAs with different methods in the literature, for example, QCD sum rule~\cite{Ioffe:1981kw,Chung:1981cc,Chernyak:1984bm,King:1986wi,Chernyak:1987nu,Braun:2000kw,Braun:2006hz}, lattice simulation~\cite{Gockeler:2008xv,Braun:2008ur,Braun:2014wpa,Bali:2015ykx,Braun:2016awp,RQCD:2019hps}, and the phenomenological models~\cite{Chernyak:1984bm,King:1986wi,Chernyak:1987nu,Bolz:1996sw,Braun:2006hz,Anikin:2013aka}. Several works are dedicated to their applications in the nucleon electromagnetic form factors, e.g., ~\cite{Braun:2001tj,Braun:2006hz,Lenz:2009ar,PassekKumericki:2008sj,Anikin:2013aka,Anikin:2014jka}. 
\par

\begin{table*}[htpb]
\centering 
\caption{List of different estimates for the parameters $f_N,\lambda_1,\varphi_{10},\varphi_{11},\eta_{10}$ and $\eta_{11}$. The values are presented at the renormalization scale $\mu^2=1$GeV$^2$ by using one-loop evolution. The errors are not shown here. The values of asymptotic (Asy.) model and QCD sum-rule (SR) estimate are taken from~\cite{Braun:2006hz}. Especially, it is well-known that the parameters here are related to the local quark-gluon operators which the QCD sum-rule methods can be applied in~\cite{Ioffe:1981kw,Chung:1981cc,Chernyak:1984bm,King:1986wi,Chernyak:1987nu,Bergmann:1999ud,Braun:2000kw,Braun:2006hz,Stefanis:1997zyh}. The Braun-Lenz-Wittman (BLW) model is also shown here~\cite{Braun:2006hz}. This model is based on the light-cone sum-rule analysis on the nucleon electromagnetic form factors which results in a relation between the electromagnetic form factors and the nucleon twist-3 and twist-4 DAs. With the choice of specific values, a good description for the experimental data was obtained. Furthermore, the parameters have been estimated in the lattice QCD simulations~\cite{Gockeler:2008xv,Braun:2008ur,Braun:2014wpa,Bali:2015ykx,Braun:2016awp,RQCD:2019hps} and the updated values evaluated by RQCD collaboration are adopted~\cite{RQCD:2019hps}. }
\label{table}
\begin{tabular}{|c|c|c|c|c|c|c|}
\hline 
 &  $10^3f_N$(GeV$^2)$ &$10^3\lambda_1$(GeV$^2)$  &$\varphi_{10}$& $\varphi_{11}$&$\eta_{10}$ & $\eta_{11}$ \\
\hline 
Asy.&  $5$ &$-27$  &0& 0&0 & 0\\ 
\hline 
RQCD&  $3.63$ &$-41.63$  &$0.198$& $0.130$&\diagbox{}{} &\diagbox{}{}  \\ 
\hline 
SR&  $5 $ &  $-27$ & 0.1725& $0.2075$
&$-0.0625$ & $-0.1849$ \\  
\hline 
BLW &  $5$ &$-27$  &0.0575& 0.0725 &$0.0500$ & $0.0326$ \\ 
\hline 
\end{tabular}
\end{table*}

\begin{figure}[htpb]
\subfigure[]{
\includegraphics[width=0.7\columnwidth]{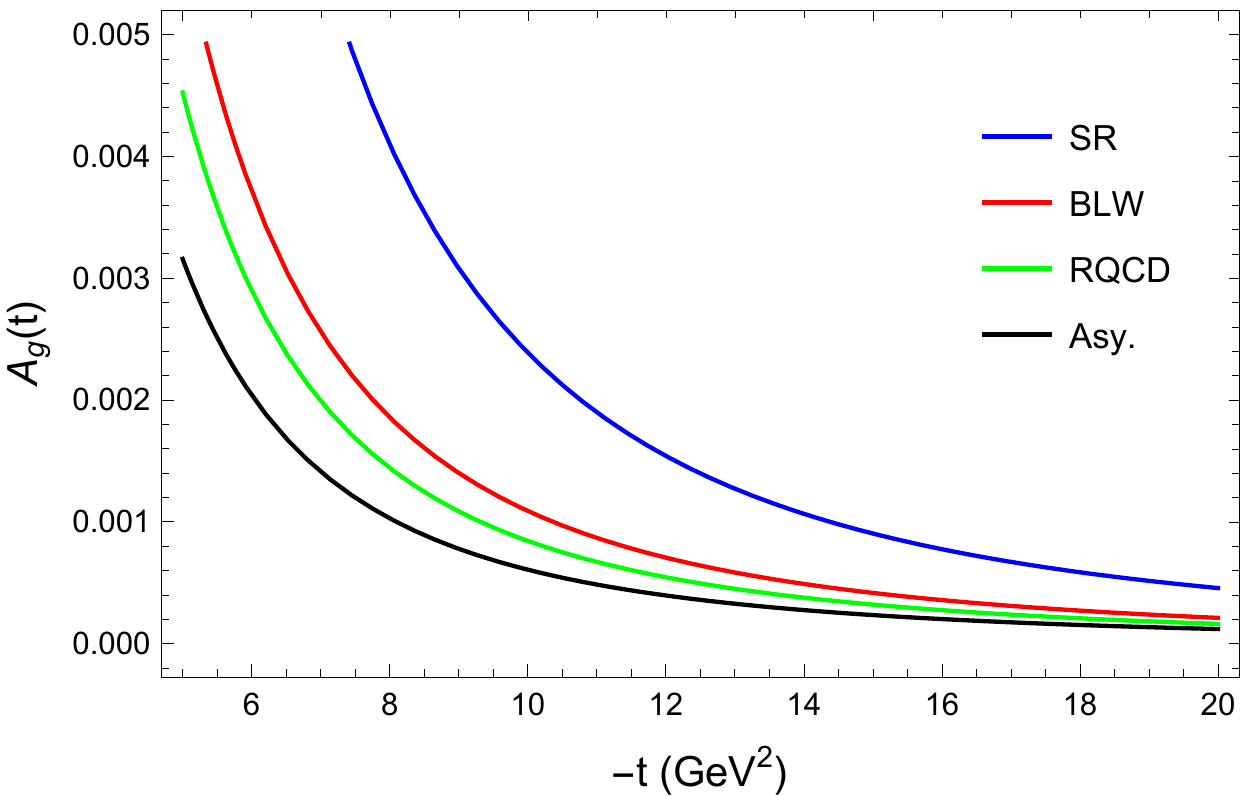}
}
\subfigure[]{
\includegraphics[width=0.7\columnwidth]{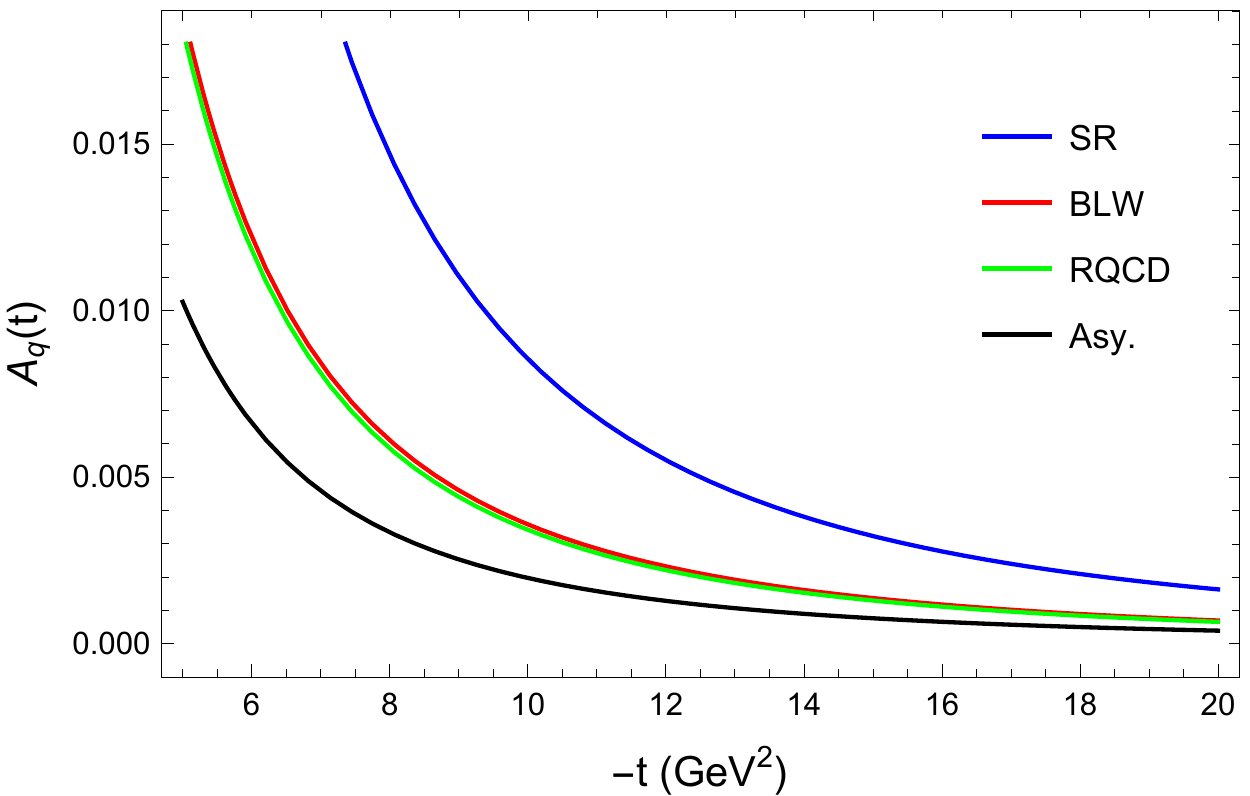}
}
\subfigure[]{
\includegraphics[width=0.7\columnwidth]{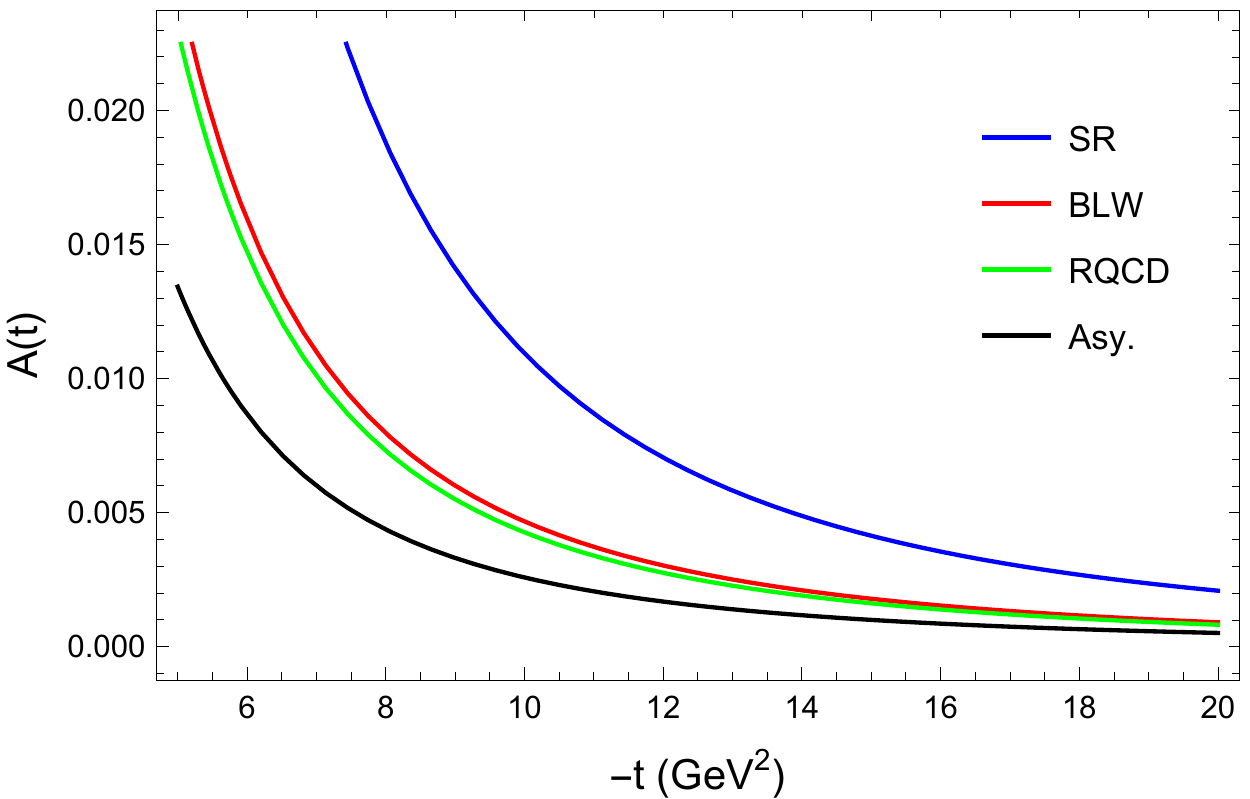}
}
\caption{$A_{q,g}(t)$ and $A(t)$ with different estimates or models of nucleon DAs: QCD sum rule(SR)~\cite{Braun:2000kw,Braun:2006hz}, Lattice simulations conducted by RQCD collaboration \cite{RQCD:2019hps}, Braun-Lenz-Wittman(BLW) model~\cite{Braun:2006hz}, the asymptotic(Asy.) model.}
     \label{fig:A1}
 \end{figure}

\begin{figure}[htpb]

\includegraphics[width=0.7\columnwidth]{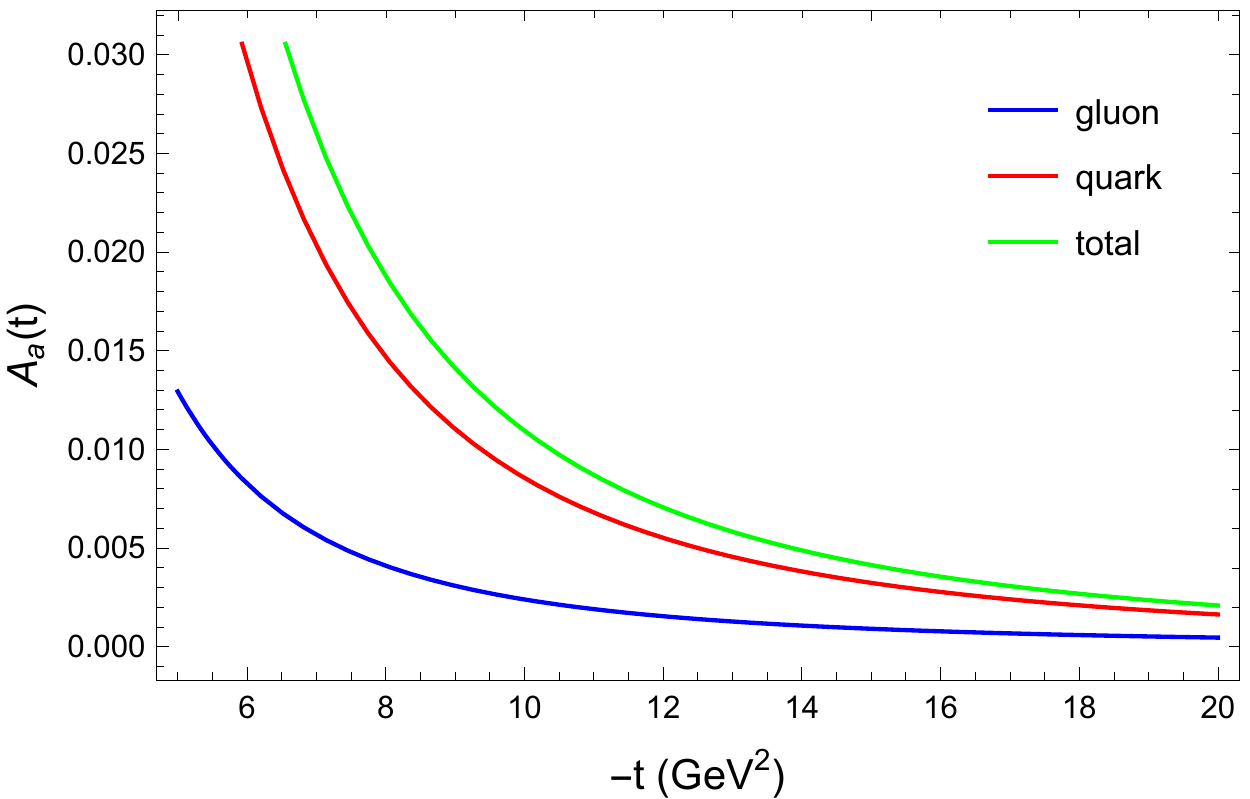}

\caption{Comparisons of $A_{q,g}(t)$ and $A(t)$ with the nucleon DAs estimated from QCD sum rule~\cite{Braun:2000kw,Braun:2006hz}.}
     \label{fig:A2}
 \end{figure}

In our calculations, we adopt the parametrization of the nucleon DAs proposed in~\cite{Braun:1999te,Braun:2008ia}.  
For the twist-3 DA, we use
 \begin{align}
  \Phi_3(x_1,x_2,x_3)= &120 x_1x_2 x_3 f_N L^{2/(3b)}
    \notag\\& \times \Big[1+ \varphi_{10}L^{20/(9b)} \mathcal{P}_{10}(x_1,x_2,x_3)
    \notag\\
  & 
 +\varphi_{11}L^{8/(3b)} \mathcal{P}_{11}(x_1,x_2,x_3)\Big]~.
  \label{protonDA1} 
 \end{align}
 where $\mathcal{P}_{10}=21\left(x_{1}-x_{3}\right),\mathcal{P}_{11}=7(1-2 x_2+ x_3)$. 
 For the twist-4 DAs, we use
\begin{align}
 \Phi_4(x_1,x_2,x_3)=& \frac{1}{2}\big[f_N L^{2/(3b)} W_\Phi(x_1,x_2,x_3)
  \notag \\
&\hspace{1.2cm}+\lambda_1 L^{-2/b}R_\Phi(x_1,x_2,x_3)\big]~,
 \notag \\
\Psi_4(x_1,x_2,x_3)=&\frac{1}{2}\big[f_N L^{2/(3b)}W_\Psi(x_1,x_2,x_3)
\notag \\ 
&\hspace{1.2cm}-\lambda_1 L^{-2/b}R_\Psi(x_1,x_2,x_3)\big]~,
  \label{protonDA2}
\end{align}
where $W_{\Phi,\Psi}$ are known as the the Wandzura-Wilczek contributions\cite{Braun:2008ia}, which can be expressed in terms of the parameters presented in the twist-3 DA:
\begin{align}
W_\Phi (x_1&,x_2,x_3)=- 40x_{1} x_{2}(2 x_3 -1 ) 
\notag \\ 
&-20\left(3-\frac{\partial}{\partial x_{3}}\right)  x_{1} x_{2} x_{3}
\Big[
\varphi_{10} L^{20/(9b)} 
 \\ 
&\times \mathcal{P}_{10}(x_1,x_2,x_3)+\varphi_{11}L^{8/(3b)} \mathcal{P}_{11}(x_1,x_2,x_3)\Big ]
~,
\notag \\ 
W_\Psi (x_1&,x_2,x_3)=- 40 x_{1} x_{3}(2 x_2 -1 ) 
\notag \\ 
&-20\left(3-\frac{\partial}{\partial x_{2}}\right)  x_{1} x_{2} x_{3}
\Big[ 
\varphi_{10} L^{20/(9b)}
  \\ 
&\times \mathcal{P}_{10}(x_2,x_1,x_3)+\varphi_{11}L^{8/(3b)} \mathcal{P}_{11}(x_2,x_1,x_3)\Big ]
~  \notag ,
\end{align}
while $R_{\Phi,\Psi}$ contain new parameters:
\begin{align}
R_\Phi(x_1&,x_2,x_3)=24x_1 x_2 \Big[1+\eta_{10}L^{20/(9b)}4(x_3+x_1-3/2 x_2)
  \notag  \\
&\hspace{1cm}-\eta_{11}L^{4/b}20/3(x_3-x_1+1/2 x_2)\Big],
 \\
R_\Psi(x_1&,x_2,x_3)=24x_1 x_3 \Big[1+\eta_{10}L^{20/(9b)}4(x_2+x_3-3/2 x_1)
\notag \\
&\hspace{1cm}+\eta_{11}L^{4/b}20/3(x_2-x_3+1/2 x_1)\Big]~.
\end{align} 
In the above equations, the scales of the parameter are hidden for brevity, e.g. $f_N\equiv f_N(\mu_0^2)$. The label $L$ stands for the one-loop evolution factor from the scale $\mu_0$ to $\mu$:
\begin{align}
L(\mu^2)=\frac{\alpha_s(\mu^2)}{\alpha_s(\mu_0^2)}.
\label{eq_L}
\end{align}
The running coupling at the leading order is 
\begin{align}
\alpha_s(\mu^2)= \frac{4\pi}{b \ln(\mu^2/\Lambda^2_{QCD})} 
\label{eq_runcouple}
\end{align}
where $b=11-2/3 n_f$ and $n_f$ is the number of the active quark flavors. Here we take $n_f=4$ and $\Lambda_{QCD}=154$ MeV. Other parameterizations are also proposed in~\cite{Braun:2000kw,Stefanis:1997zyh,Bergmann:1999ud}. 
\par 
To our accuracy, there are total of 6 non-perturbative parameters: $f_N,\lambda_1,\varphi_{10},\varphi_{11},\eta_{10}$ and $\eta_{11}$. $f_N,\lambda_1$ are the normalizations of the nucleon DAs~\cite{Braun:2008ia}. $\varphi_{10},\varphi_{11},\eta_{10}$ and $\eta_{11}$ are the shape parameters which describe the shape of the amplitudes and also determine the deviation from the asymptotic forms. Those parameters can be estimated by using the non-perturbative methods~\cite{Ioffe:1981kw,Chung:1981cc,Chernyak:1984bm,King:1986wi,Chernyak:1987nu,Braun:2000kw,Braun:2006hz,Gockeler:2008xv,Braun:2008ur,Braun:2014wpa,Bali:2015ykx,Braun:2016awp,RQCD:2019hps}. Phenomenological models have also been built from fitting to the experimental data~\cite{Chernyak:1984bm,King:1986wi,Chernyak:1987nu,Bolz:1996sw,Braun:2006hz,Anikin:2013aka}. For the detailed summary and comparison of different estimates, we refer the readers to~\cite{Stefanis:1997zyh,Lenz:2009ar,Anikin:2013aka,Braun:2014wpa,RQCD:2019hps} and the references therein. In Table~\ref{table}, we list those applied in this paper.

 \par
With the above parametrization, we can carry out the convolution integral in the factorization formula for the GFFs. The calculation is straightforward for the $A_{q,g}$ form factor. However, as we mentioned above, for the $B_{q,g},C_{q,g},\overline {C}_{q,g}$ form factors from the helicity-flip amplitude, there are end-point singularities in the convolution integral at $x_i,y_i\rightarrow0$. To regulate these divergences, we introduce a cut-off $\Lambda^2_c/(-t)$ for the integration as proposed in~\cite{Belitsky:2002kj} for Pauli form factor, where $\Lambda_c$ is a soft scale. With the leading-order results derived here, one can check explicitly that the end-point behaviors of integrand can only follow the form: $\frac{1}{x_i}, \frac{1}{y_i}$ or $\frac{1}{x_i y_j}$. Contributions from the first two forms will give the single logarithm term as $\ln(-t/\Lambda^2_c)$ and the last one would yield the double logarithm term $\ln^2 (-t/\Lambda^2_c)$. With that improvement, we find that the $B_{q,g},C_{q,g}$ scale as $\ln^2(-t/\Lambda^2_c)/(-t)^3$ at large momentum transfer, same as their total GFFs. $\overline{C}_{q,g}$ scales as $\ln^2(-t/\Lambda^2_c)/(-t)^2$. 
Similar double logarithms were observed in the pQCD analysis of Pauli form factor $F_2$ and play an important role in describing the scaling behavior of $Q^2F_2/F_1$ at the large momentum transfer as compared to the experimental data~\cite{Belitsky:2002kj,Puckett:2010ac}. 
 \par 

In Fig.~\ref{fig:A1} we show the $(-t)$-dependence of the GFFs $A_{q,g}(t)$ and $A(t)=A_{q}(t)+A_g(t)$ obtained from different estimates of twist-3 nucleon DAs. Several features can be found. First, it is easy to see that all $A$-form factors are positive, which show the same sign with the lattice simulations of GFFs~\cite{Hagler:2007xi,Shanahan:2018pib} at low $-t$. Second, different models differ significantly at low $(-t)$. However, at higher $(-t)$, the QCD evolution tends to reduce the model dependence for the distribution amplitudes and all predictions are within the same order of magnitude. Third, among the models, the $A$-form factor from the asymptotic DA has the smallest magnitude over the whole range of $-t$, while the DAs obtained from the QCD sum-rule yield the largest magnitude. Moreover, the numerical evaluations show the following relations between the $A$-form factors: $A(t)>A_q(t)>A_g(t)$. For example, in Fig.~\ref{fig:A2}, we present the comparisons of $A_{q,g}(t)$ and $A(t)$ with the DAs obtained from the QCD sum rule. Similar relations can be observed in the previous lattice calculations of GFFs in the range $0.3$GeV$^2<-t<2$GeV$^2$~\cite{Shanahan:2018pib,Hagler:2007xi}.
\par

\begin{figure}[htbp]
\subfigure[]{
\includegraphics[width=0.7\columnwidth]{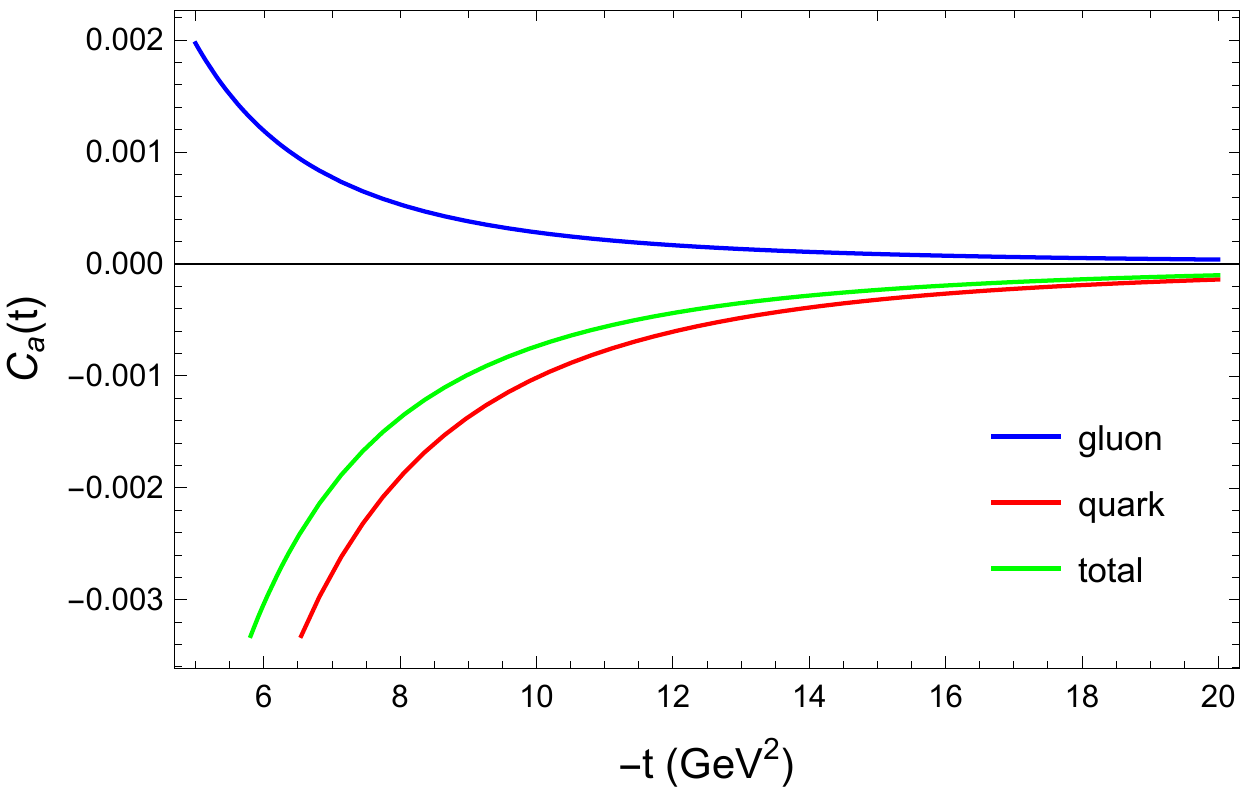}
     }
     \subfigure[]{
\includegraphics[width=0.7\columnwidth]{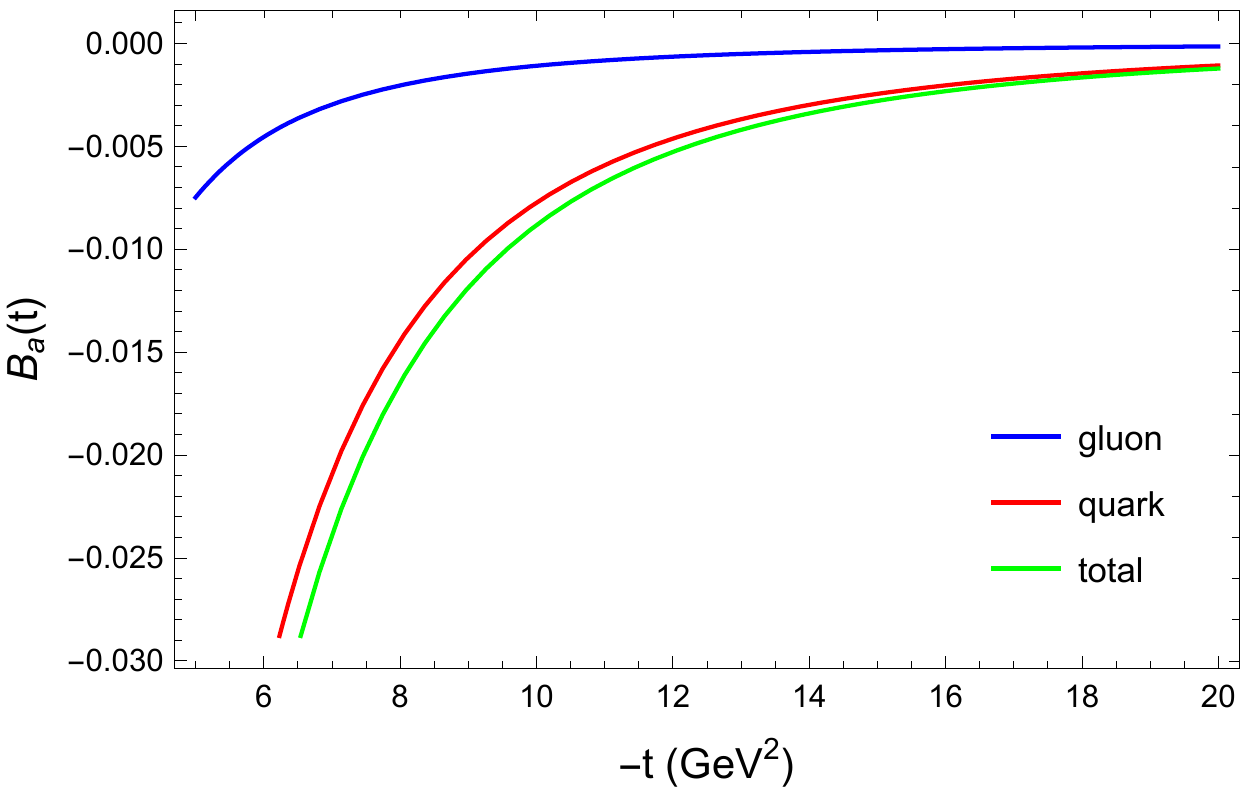}
 }

\subfigure[]{
\includegraphics[width=0.7\columnwidth]{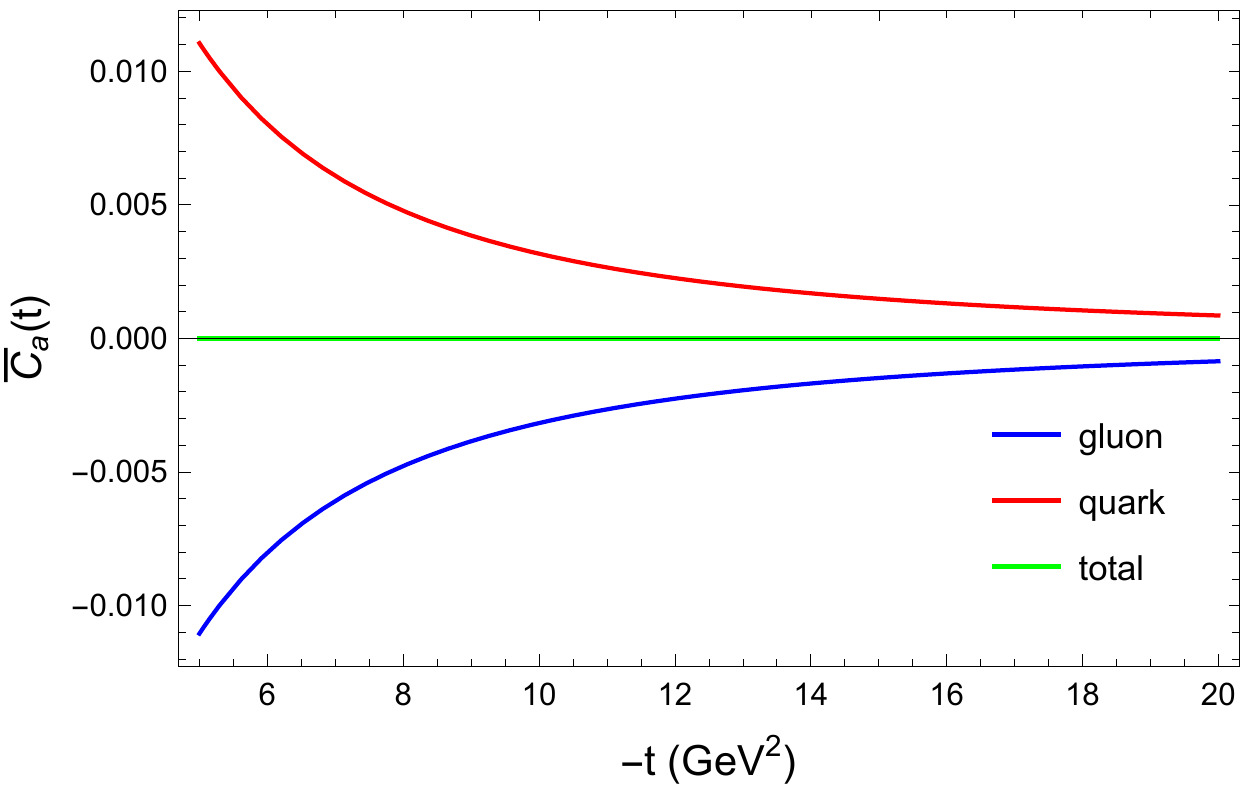}
 }
\caption{Nucleon GFFs with nucleon DAs estimated from QCD sum rule~\cite{Braun:2000kw,Braun:2006hz}. Here we take the soft scale $\Lambda_c=200$MeV to regulate the end-point singularities associated with the nucleon helicity flip amplitude to extract the GFFs.}
     \label{fig:GFFsSR}
 \end{figure}

  \begin{figure}[htbp]
\includegraphics[width=0.7\columnwidth]{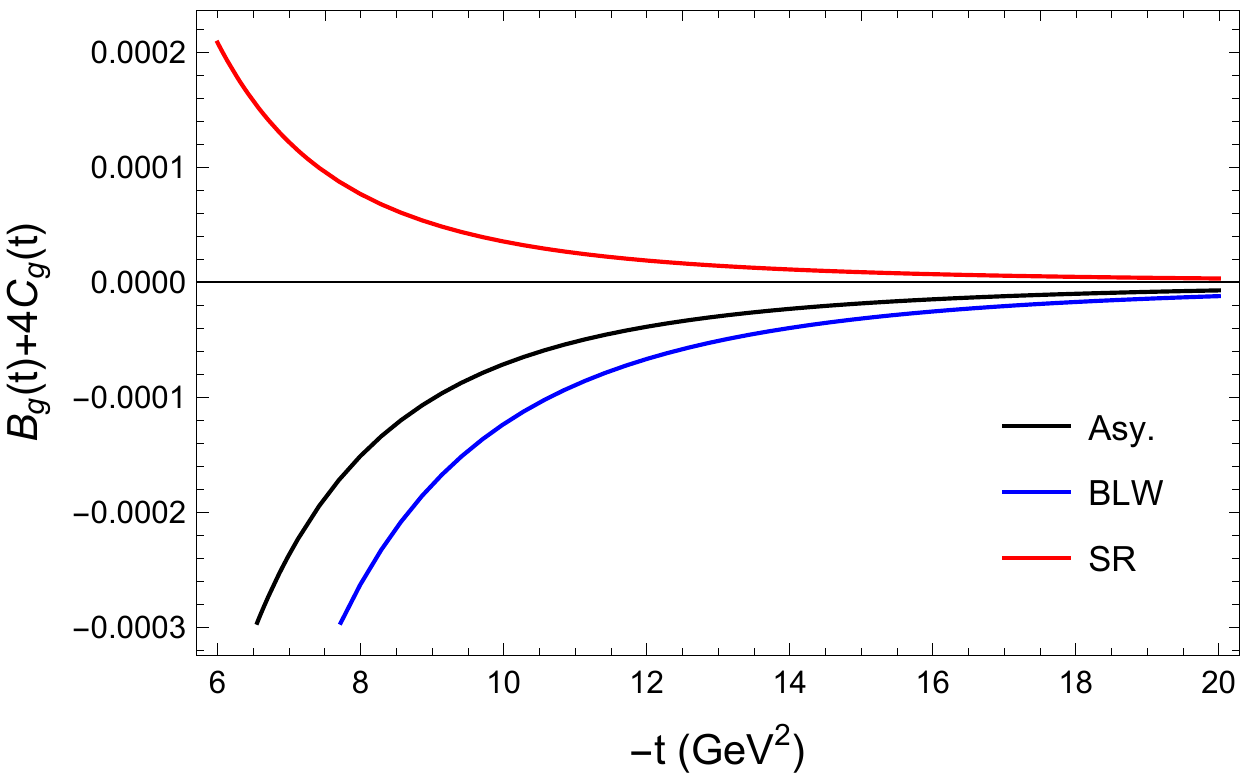}
\caption{$B_{q}(t)+4C_g(t)$ with different estimates or models of nucleon DAs:  QCD sum rule(SR)~\cite{Braun:2000kw,Braun:2006hz}, Braun-Lenz-Wittman(BLW) model~\cite{Braun:2006hz}, the asymptotic(Asy.) model.}
    \label{fig:CB}
 \end{figure}

  \begin{figure*}[htbp]
\subfigure[]{
\includegraphics[width=0.7\columnwidth]{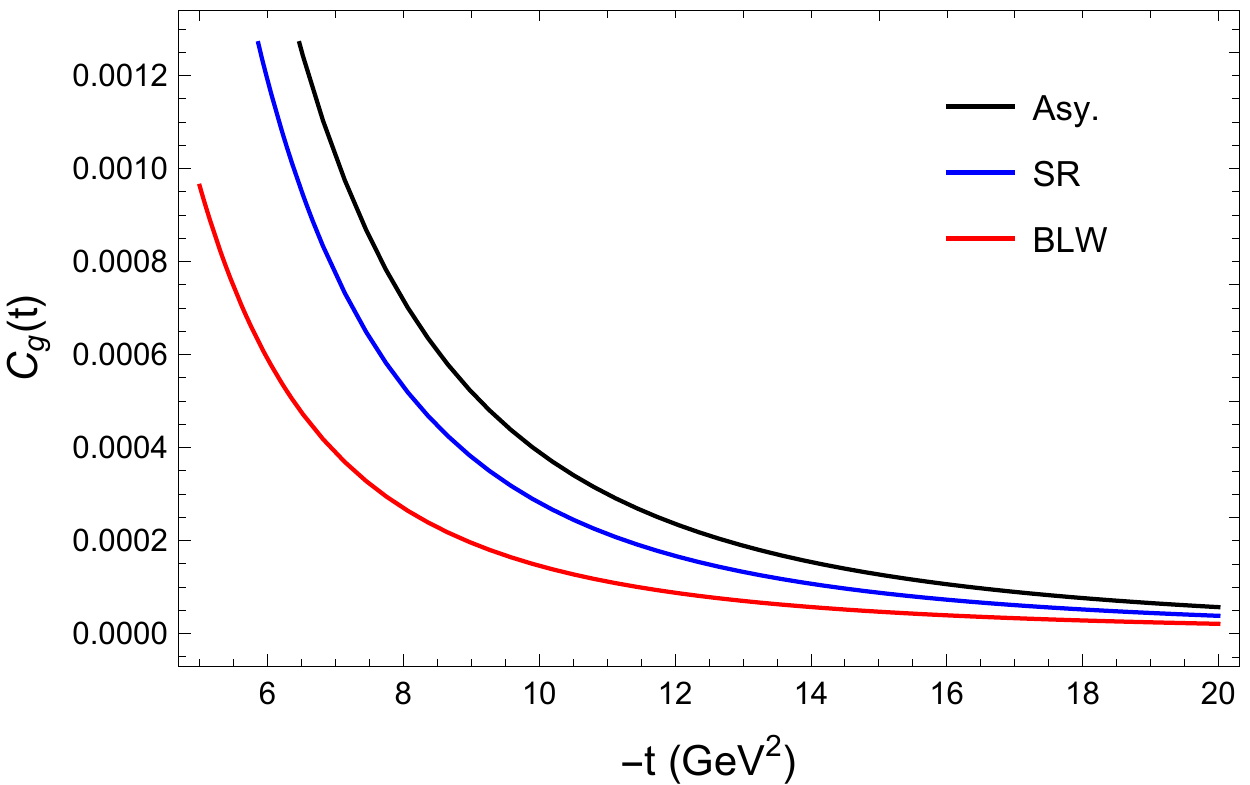}
}
\subfigure[]{
\includegraphics[width=0.7\columnwidth]{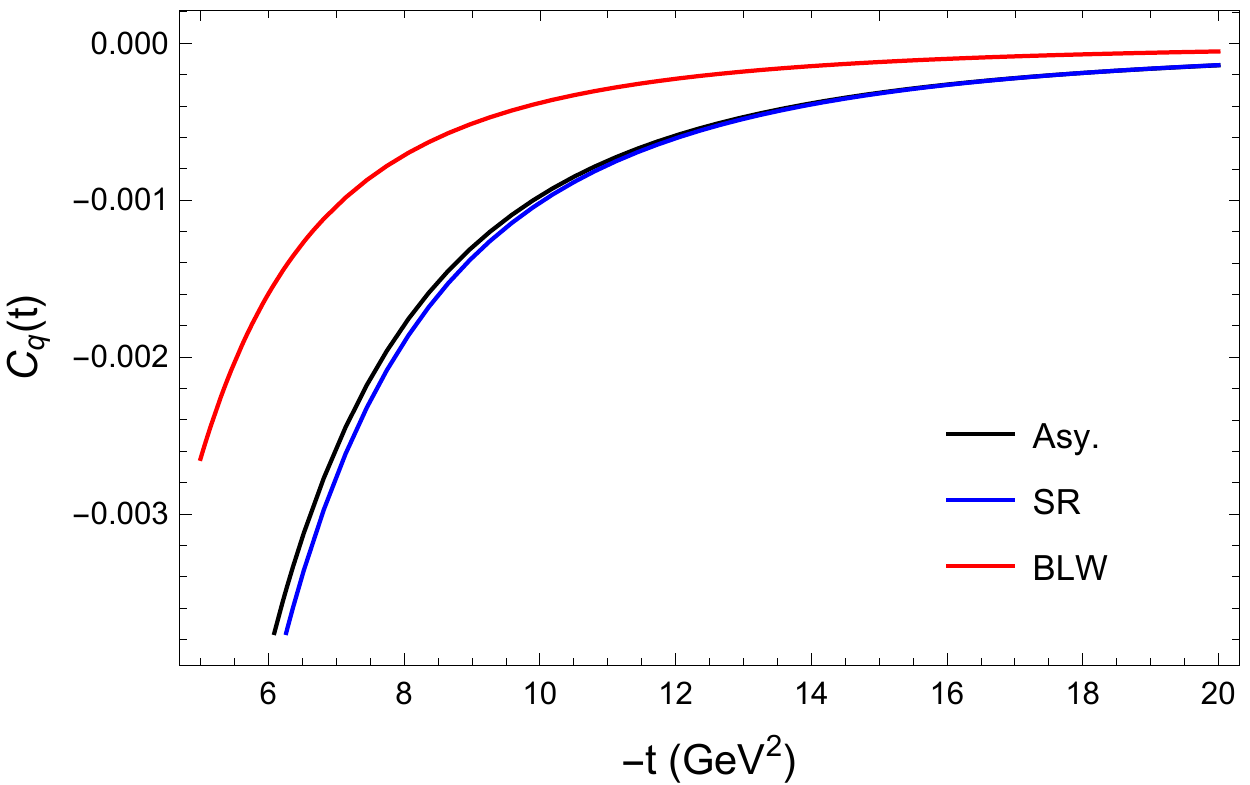}
}

\subfigure[]{
\includegraphics[width=0.7\columnwidth]{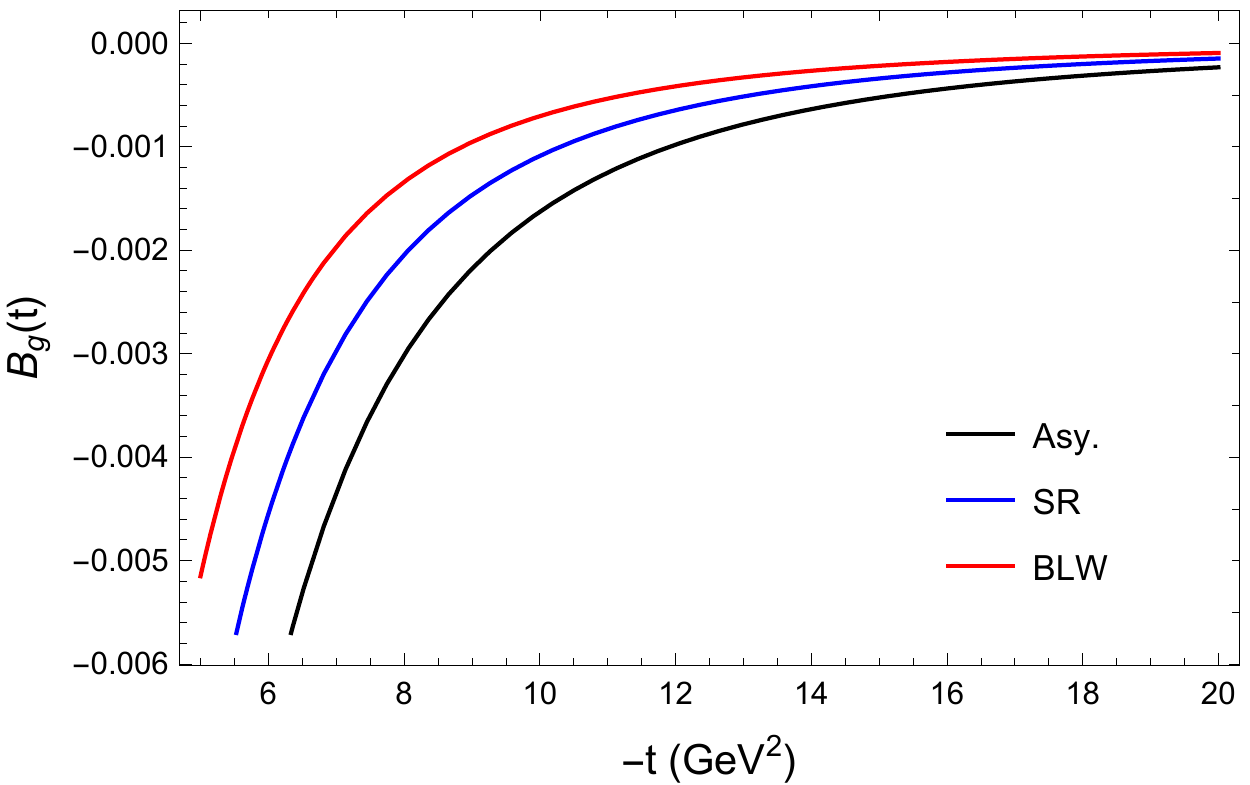}
}
\subfigure[]{
\includegraphics[width=0.7\columnwidth]{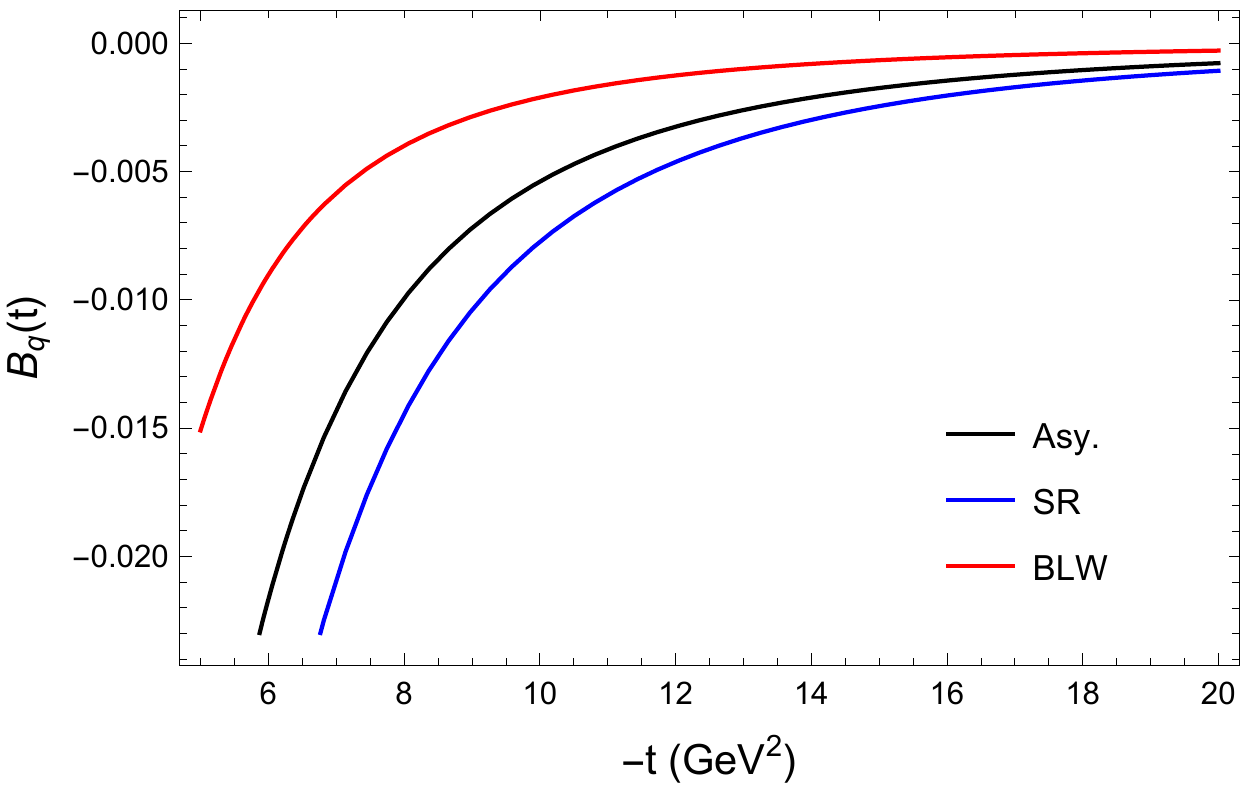}
 }
\subfigure[]{
\includegraphics[width=0.7\columnwidth]{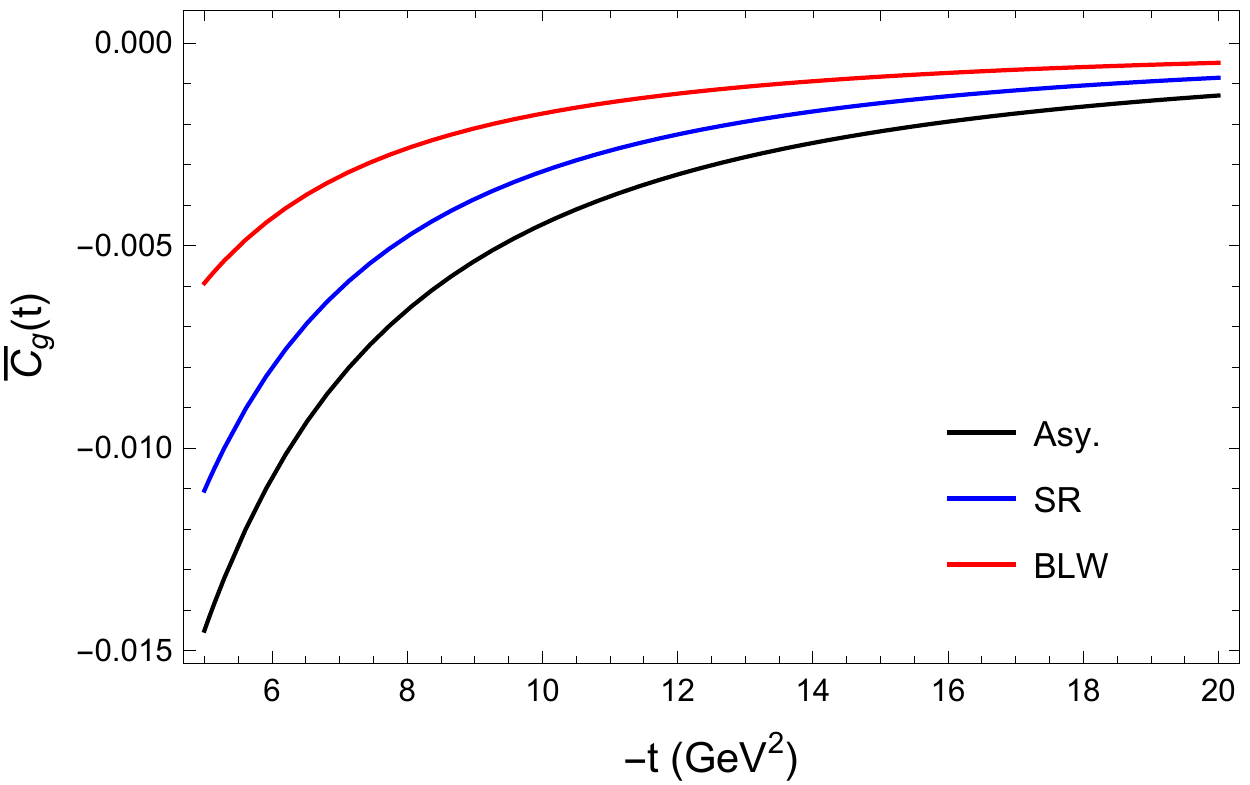}
}
     \caption{Nucleon GFFs with different estimates or models of nucleon DAs:   QCD sum rule(SR)~\cite{Braun:2000kw,Braun:2006hz}, Braun-Lenz-Wittman(BLW) model~\cite{Braun:2006hz}, the asymptotic (Asy.) model. Here we take $\Lambda_c=200$MeV.}
     \label{fig:GFFsModel}
 \end{figure*}

\par 
In Figs.~\ref{fig:GFFsSR}-\ref{fig:GFFsModel}, we show the numeric results for $B$, $C$ and $\overline C$ form factors. These GFFs are extracted from the nucleon helicity-flip amplitudes. To regulate the end-point singularities, we introduce a soft scale $\Lambda_c$. 
First, we can make the comparisons among the quark, gluon and total GFFs with a fixed $\Lambda_c$. Some general features can be found for the models in Table~\ref{table}. For the $C$-form factors, $C_q(t)$ and $C(t)$ are negative while $C_g(t)$ is positive, and they have the relations $|C_q(t)|>|C(t)|>|C_g(t)|$. On the other hand, the lattice calculation in \cite{Shanahan:2018pib} shows that $C_g(t)$ is negative at low $-t$. That means the $C_g(t)$ will change sign at higher $(-t)$. 
For the $B$-form factors, they are all negative with the relations $|B_g(t)|<|B_q(t)|<|B(t)|$. For the $\overline C$-form factors, the total $\overline{C}$-form factor vanishes as we confirmed explicitly before and $\overline{C}_q(t)=-\overline{C}_g(t)>0$. In particular, we display the above comparisons among the GFFs with the nucleon DAs from QCD sum rule~\cite{Braun:2000kw,Braun:2006hz} in Fig.~\ref{fig:GFFsSR} where $\Lambda_c$ is taken as $200$MeV.

\par 
On the other hand, from our perturbative results at leading order, one can check explicitly that the combination $B_g(t)+4C_g(t)$ is indeed free from end-point singularity and hence do not need the cut-off scale $\Lambda_c$. In Fig.~\ref{fig:CB}, we show the $(-t)$-dependence of $B_g(t)+4C_g(t)$ obtained from different modeling of DAs. Numerically, they are also order of magnitudes smaller compared to the results shown in Figs.~\ref{fig:GFFsSR}.

\section{Scalar Form Factors \label{sec:SFF} }

The scalar form factors(FFs), defined as the hadron form factors of the $F^2$ operator, are also important in GFFs physics, since it has the close connection to the trace anomaly, as shown in Eq.~(\ref{eq_traceanomaly}). Following the discussion in the previous sections, similar perturbative analysis can be applied to the scalar FFs at large $-t$. We will first study the pion case, which will be simple and interesting, and then turn back to the proton case.  

\subsection{Pion \label{sec:scalarFF_pion}}

The scalar FF for pion, $G_\pi(t)$, is defined from the parameterization of the $F^2$ transition matrix: 
 \begin{align}
\langle P'|F^a_{  \mu \nu }(0) F^{a\mu \nu} (0)|P\rangle= m_\pi^2 G_\pi(t)~.
\end{align}
At large $(-t)$, the factorization analysis can be applied. 
At the leading order of $\alpha_s$, there is only one diagram that contributes, see, for example, Ref.~\cite{Tong:2021ctu}. Therefore,
\begin{align}
m_\pi^2& G_\pi(t)
\notag \\
=& \int d xd y \ \phi^*(y)\phi(x)\frac{\text{tr}[t^a t^a]}{3}
\ 
\frac{g_s^2\text{Tr}[ \slashed P'\gamma_\sigma  \slashed P\gamma_\rho]}{4(q_1^2+i \epsilon)(q_2^2+i \epsilon)}  
\notag \\ &  
\Big[(-q_{1}^\mu g^{\nu \sigma}+q_1^\nu g^{ \mu \sigma })(-q_{2\mu} g_\nu^{\ \rho}+q_{2\nu} g_\mu^{\ \rho} )+(\sigma \leftrightarrow \rho)\Big]
\notag\\
=&\int d xd y\  \phi^*(y)\phi(x)\ 8\pi \alpha_s C_F \left (\frac{1}{x \bar y}+\frac{1}{\bar x  y} \right )+{\cal O}(t^{-1})~,
\label{eq_pionscalar1}
\end{align}
where $q_1=x P-y P',\ q_2=\bar x P-\bar y P'$ are the momenta that flow into the $F^2$ operator. $\phi(x)$ is the pion twist-2 light-cone distribution amplitude, defined as 
\begin{align}
&\phi(x)i \sqrt{6}P^+
\notag \\ 
=& \int  \frac{d \xi^-}{2\pi} e^{i x P^+ \xi^-}\langle 0|\bar{d}(0)[0, \xi^-] \gamma^+ \gamma_{5} u( \xi^-)| \pi(P)\rangle~,
\end{align}
with the normalization
$
\int^1_0 dx \  \phi(x)= \frac{f_\pi}{\sqrt{6}},
$
where $f_\pi\approx92.3$MeV is the pion decay constant. 
\par 
From the above results, it is interesting to see that the pion scalar form factor has no power behavior with respect to $-t$, which is different from the pion $A,C$ form factor with $A^\pi_{g,q}\sim C^\pi_{g,q}\sim 1/(-t)$~\cite{Tong:2021ctu}. 
It is also expected that higher order contributions will not change the power behavior. 
At leading order perturbation theory, the possible $(-t)$-dependence only comes from the running of the strong coupling $\alpha_s$ with $\mu^2$ taken as $-t$.
There are also evolution effects hidden in the pion DAs. Hence, to make a sensible numerical estimate on the large $-t$ behavior of the pion scalar form factor, more details on the pion DAs are needed.
\par
Similar to the proton DAs introduced in the previous sections, the pion DA can also be expressed as the series in the the eigenfunctions of the leading order evolution equation (e.g.~\cite{Fagundes:2018dyn}):
\begin{align}
\phi(x,\mu^2)=&\sqrt{6} f_\pi x \bar x \Big [1+ 
\sum_{n=1}^\infty a_{2n}(\mu^2_0) L^{\gamma^{(0)}_{2n}/b}_{2n}(\mu^2) C_{2n}^{3/2}(x-\bar x)\Big ]~,
\label{eq_pionDAGegen}
\end{align}
where $C_n^\alpha(x)$ are the Gegenbauer polynomials. $L_n(\mu^2)$ has been given in Eq.~(\ref{eq_L}) and represents the logarithmic factor for the one-loop evolution from the default scale $\mu_0^2$ to the varied scale $\mu^2$. $\gamma_{n}$ is  the anomalous dimension given by 
\begin{align}
\gamma^{(0)}_{n}=4 C_{F}\left(\psi(n+2)+\gamma_{E}-\frac{3}{4}-\frac{1}{2(n+1)(n+2)}\right)~,
\end{align}
where $\psi(n)$ is the digamma function and $b=11-2/3 n_f$. When $\mu^2$ approaches to $+\infty$, all the logarithmic factors are suppressed, and the first term determines the asymptotic form of the DA, i.e., $\phi^{asy}(x)=\sqrt{6} f_\pi x(1-x)$. The coefficients $a_{2n}$, known as Gegenbauer moments, describe the deviation from the asymptotic DA. 
\par 
Following Eq.~(\ref{eq_traceanomaly}), it is convenient to express the pion transition matrix element of the trace anomaly in terms of the pion scalar form factor in massless-quark limit.  With the Gegenbauer expansion of the pion DAs in Eq.~(\ref{eq_pionscalar1}) and $\mu^2=-t$, a simple expression can be obtained   
\begin{align}
\langle P'|
\frac{\beta(g)}{2g}F^2|P\rangle
=&-3
\alpha_s^2(-t) C_F\left (11 -\frac{2n_f}{3} \right)  f_\pi^2  
\notag \\ &\times \Big[1+\sum_{n=1}^\infty a_{2n}(\mu_0^2) L^{\gamma_{2n}/b}_{2n}(-t)\Big ]^2 \ ,
\end{align}
where only the sum of the Gegenbauer moments is relevant. The above result shows that the pion trace anomaly matrix element should be negative in the large $-t$ limit. However, it is known that at zero momentum transfer, the pion trace anomaly should be positive~\cite{Ji:1994av,He:2021bof}. 
The above results reveal the fact that the trace anomaly matrix element should change sign from the large $(-t)$ to small $(-t)$.
Preliminary lattice simulation also indicates this behavior~\cite{KFLiu2022}. Sign changes have also been observed in the gluon $C$ form factor for both pion and proton. Later we will also show the proton scalar form factor experiences the sign change. 
\par 
With a truncation on the Gegenbauer-polynomial expansions and the explicit setting on the values of the coefficients $a_{2n}$ at the default scale $\mu_0^2$, several models on the pion DA can be obtained. Here we follow the models adopted in~\cite{Gao:2021iqq}.  In Model I~\cite{Cheng:2020vwr}, the coefficients are set as
\begin{align}
\left\{a_{2}, a_{4},a_{6}, a_{8}\right\}\left(\mu_{0}^2\right)\approx\{0.269, 0.185,0.141, 0.049\}~,
\label{eq_pionModelI}
\end{align}
where $\mu_0=1$GeV and the uncertainties are neglected here. These values are fitted from the QCD light-cone-sum-rule results for the pion electromagnetic form factor to its measured data~\cite{Cheng:2020vwr} . In model II~\cite{Bakulev:2001pa,Mikhailov:2016klg,Stefanis:2020rnd}, the coefficients are set as 
\begin{align}
\left\{a_{2}, a_{4}\right\}\left(\mu_{0}^2\right)\approx\left\{0.203,-0.143\right\}
\label{eq_pionModelII}
\end{align}
with $\mu_0=1$GeV. The values are taken from an estimation from QCD sum rules with nonlocal condensate~\cite{Mikhailov:1991pt,Bakulev:2001pa,Mikhailov:2016klg,Stefanis:2020rnd}. Besides the Gegenbauer-type models, the Ads/QCD inspired model(Model III) are also applied~\cite{Brodsky:2007hb,Gao:2021iqq}:
\begin{align}
\phi\left(x, \mu^2_{0}\right)= \frac{f_\pi}{\sqrt{2N_c}}\frac{\Gamma\left(2+2 \alpha_{\pi}\right)}{\Gamma^{2}\left(1+\alpha_{\pi}\right)}(x \bar{x})^{\alpha_{\pi}}
\label{eq_pionModelIII}
\end{align}
with $\alpha_{\pi}\left(\mu_{0}\right)\approx 0.422$. 

  \begin{figure}[htbp]
\includegraphics[width=0.7\columnwidth]{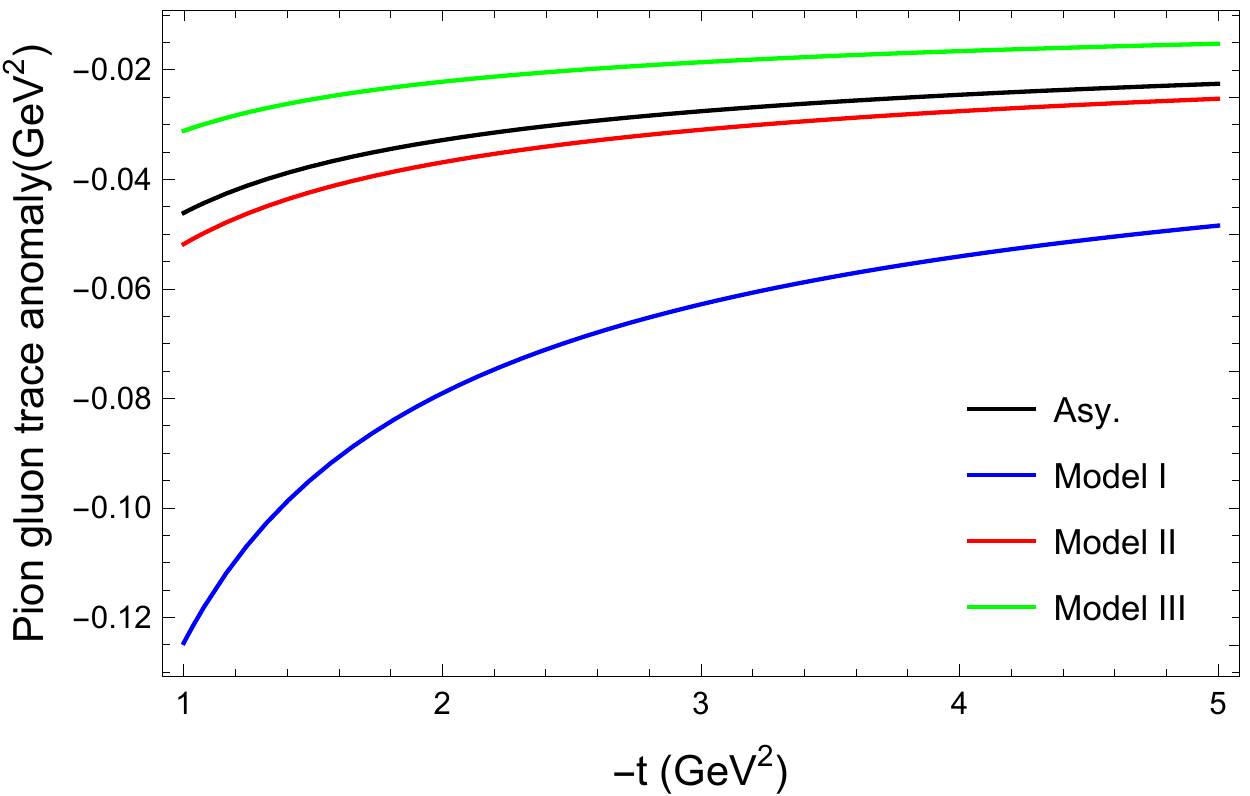}
\caption{The scalar form factor for pion $\langle P' | 
\frac{\beta(g)}{2g}F^2|P\rangle$ as function of momentum transfer squared $-t$. 
}
    \label{fig:piontraceAnomaly}
 \end{figure}

Adopting the above models, we conduct the numerical evaluations on the pion gluon trace anomaly transition matrix. The results are displayed in Fig.~\ref{fig:piontraceAnomaly} with respect to $(-t)$ in the range from 1 GeV$^2$ to 5 GeV$^2$. Here we take $n_f=4$ and $\Lambda_{QCD}=154$MeV.

\subsection{Nucleon \label{sec:scalarFF_nucleon}}

Now we turn to the nucleon case. The nucleon scalar form factor $G_p(t)$ is defined by 
\begin{align}
\langle P',s' |F^a_{  \mu \nu }(0) F^{a\mu \nu} (0)|P,s\rangle
= G_p(t)M\bar u_{s'}(P') u_s(P)~,
\end{align}
where $M$ is the nucleon mass and $u_s(P)$ is the Dirac spinor with spin $s$. From the Dirac structure, $G_p(t)$ is relevant to the nucleon helicity-flip amplitude at the high energy. Hence, we can apply the same methodology introduced in Sec.~\ref{sec_flip} to investigate the large momentum transfer behavior of $G_p(t)$. It results in a factorization formula as follows:
\begin{align} 
G_p(t)=& \int [d x][d y]\{ x_3 \Phi_4(x_1,x_2,x_3){\cal G}_{\Phi }(\{x\},\{y\})
\notag \\ &
+x_1\Psi_4(x_2,x_1,x_3) {\cal G}_{\Psi }(\{x\},\{y\})
\} \Phi_3(y_1,y_2,y_3)~,
\end{align}
where the hard function has the following structure
\begin{align}
{\cal G}=2 {\cal H}+{\cal H}'
\end{align}
and $H'=H(y_1\leftrightarrow y_3)$.
Explicitly, we have
\begin{align}
{\cal H}_\Psi=& \frac{  C_B^2 M _p  }{6t^2} (4\pi \alpha_s)^2
 \Big[ 
 x_3 \left(\left(y_1-y_3\right) \bar{x}_1+x_1 y_2\right) T_1 
 - T_3
 \notag \\ 
 &+  \bar{x}_3 \left(y_3 \bar{x}_3+x_3 \bar{y}_3\right)\tilde T_1
+  x_3\left(y_2 \bar{x}_2+x_2
   \bar{y}_2\right)(\tilde T_2-T_2)
\notag \\ 
&+
 x_3
   \left(y_1-\bar y_1\right)(T_4 + T_5)+
\left(y_3 \bar{x}_3+x_3 \bar{y}_3\right) (\tilde T_4+\tilde T_5)
\Big]~,
    \notag \\
 \end{align}
where ${\cal H}_\Phi= {\cal H}_\Psi(1\leftrightarrow 3)$ and the functions $T_i$ are defined by \begin{align}
&T_1=\frac{1}{x_1 x_3^2 y_1 y_3^2 \bar{x}_1^2 \bar{y}_1}~,
\quad
T_2=\frac{1}{x_1 x_2 x_3^2 y_2 y_3^2 \bar{x}_2 \bar{y}_2}~,
\quad
 \notag \\
&T_3=\frac{1}{x_1 x_2 y_1 y_2 \bar{x}_1^2
   \bar{y}_1}~,
\quad
T_4=\frac{1}{x_1 x_3^2 y_1 y_3
   \bar{x}_1 \bar{y}_1^2}~,
    \notag \\
 &T_5=\frac{1}{x_1 x_2 x_3 y_1 y_2
   \bar{x}_1 \bar{y}_1^2}~,\quad \tilde{T}_i=T_i(1\leftrightarrow3)~.
\end{align}
The above calculation reveals that the nucleon scalar form factor has the same power behavior as $\overline{C}_{g,q}$, i.e., $G_p(t)\sim1/(-t)^2$ and hence $\langle P'_\uparrow|F^2|P_\downarrow\rangle\sim 1/(-t)^{3/2}$. The convolution integral also suffers from the end point singularities, and the logarithm terms are expected.

  \begin{figure}[htbp]
\includegraphics[width=0.7\columnwidth]{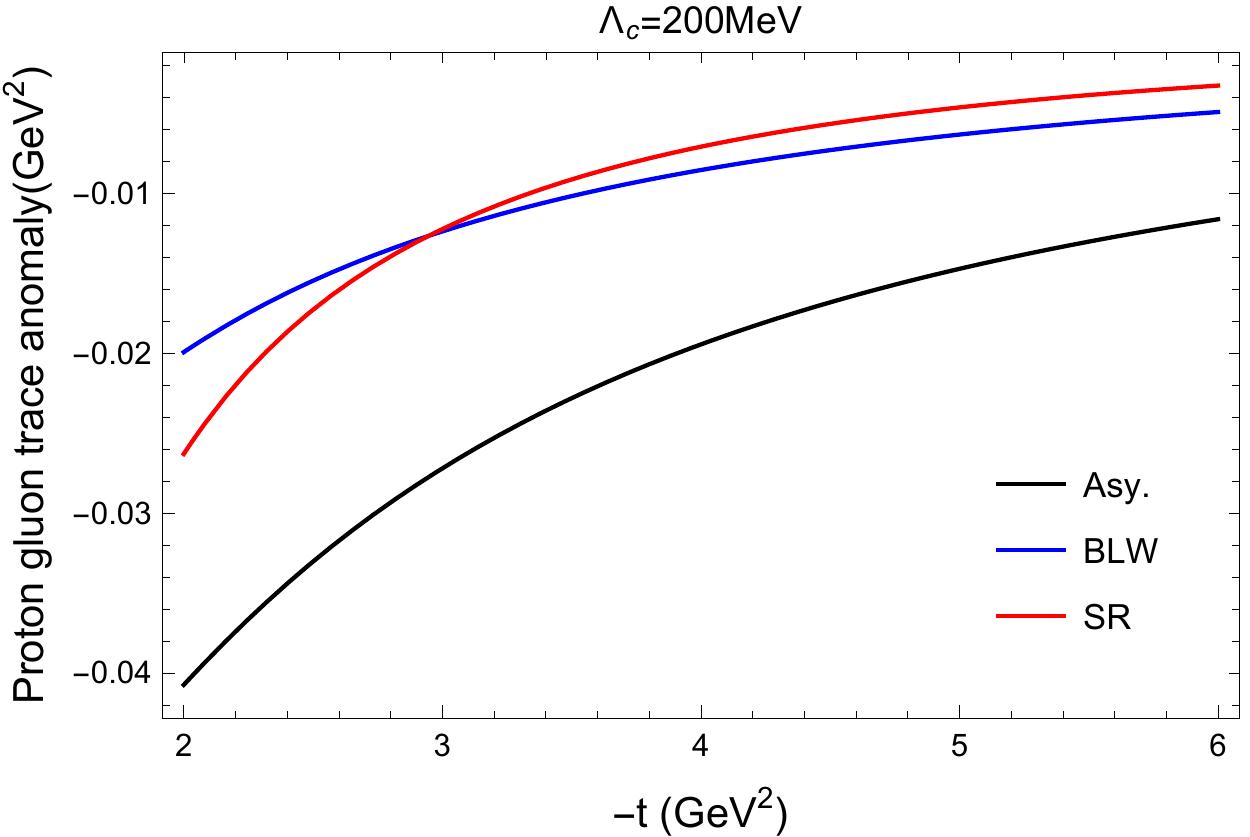}
\caption{ $\langle P'_\uparrow | \frac{\beta(g)}{2g}F^2|P_\downarrow\rangle$ with respect to $-t$ within different models of nucleon DAs. 
}
    \label{fig:piontraceAnomaly1}
 \end{figure}

With the above factorizations formula, we apply the methods introduced in Sec.~\ref{sec:numerical} to make numeric estimates on the gluon trace anomaly transition matrix for the nucleon, $\langle P'_\uparrow|\frac{\beta(g)}{2g}F^2|P_\downarrow\rangle $. Similarly, the soft scale $\Lambda_c$ is introduced to regulate the end point singularities that appear in the hard part. In Fig.~\ref{fig:piontraceAnomaly1}, we present the result in the range $0.35$ GeV$^2 <|t|<5$GeV$^2$ with three models of nucleon DAs in Table \ref{table}, where $\Lambda_c=200$ MeV, $n_f=4$ and $\Lambda_{QCD}=154$ MeV. It shows that the nucleon gluon trace anomaly transition matrix element is negative in the large$-|t|$ region. On the other hand, it is known that at zero momentum transfer, it yields the proton quantum anomalous energy and is positive~\cite{Ji:1994av}. That means that the transition matrix element should change sign when $|t|$ varies from the low$-|t|$ region to the high$-|t|$ region. By inspecting the results obtained from the explicit models of nucleon DAs, one can see the sign change comes from the end point contributions, for example, in the asymptotic-DA model,
\begin{align}
\langle P'_\uparrow | \frac{\beta(g)}{2g}F^2|P_\downarrow\rangle\bigg\vert_{\text{Asy.} } =&-\frac{49168f_N^2 \alpha _s^3(-t)}{(-t)^{3/2}} \Bigg [\log ^2\left(\frac{\Lambda_c
   ^2}{-t}\right)
   \notag \\ 
   &+4.5 \log \left(\frac{\Lambda_c
   ^2}{-t}\right)+4.75\Bigg]~,
\end{align}
where the constant term and the double-logarithm term in the square bracket have negative contributions to the transition matrix, while the single logarithm has positive contributions and dominate in the low-$|t|$ region. Although the numerical results above are based on the perturbative analysis and in principle, only make sense in the large momentum transfer region($|t|\gg \Lambda_{QCD}^2$),  we expect  the soft end-point contributions should be one of the origins of the sign change. 

\section{Conclusion}
In this paper, we have performed a detailed perturbative analysis of the gravitation form factors for nucleon at large momentum transfer, which results in the factorizations formulas of the GFFs in terms of the twist-3 or twist-4 nucleon distribution amplitude. We derived the hard coefficients at the leading order explicitly. 

For $A_{q,g}$ form factors, we have demonstrated that they are related to the helicity-conserved amplitude and the quark and gluon contributions both scale $1/(-t)^2$. We have also computed $B_{q,g},C_{q,g},\overline C_{q,g}$ form factors separately from the helicity-flip amplitude. Multiple consistent checks have been performed. We found that $C_{q,g},B_{g,q}$ are power suppressed as compared to the $A_{q,g}^p$. $\overline C_{q,g}$ scale as $A_{q,g}^p$ due to parameterization and the total $\overline C$ vanishes as expected. Because of the end point singularities, the $B_{q,g},C_{q,g},\overline C_{q,g}$ form factors receive additional logarithmic contributions.

We have also computed the scalar form factors $\langle P'|\frac{\beta(g)}{2g}F^2|P\rangle$ for pion and proton and found out that both are negative at large $(-t)$. It is understood that they are related to the hadron masses at zero momentum transfer. Therefore, there must be a sign change at lower $(-t)$. This may lead to a nontrivial mass distributions of hadrons.

Based on the models for the twist-3 and twist-4 distribution amplitudes, we have performed numeric estimates for the quark and gluon GFFs in the perturbative region of large momentum transfer. It will be interesting to test these predictions in future experiments. 

{\bf Acknowledgments:} We thank Xiangdong Ji, Keh-Fei Liu for discussions and comments. This material is based upon work supported by the LDRD program of LBNL, the U.S. Department of Energy, Office of Science, Office of Nuclear Physics, under contract numbers DE-AC02-05CH11231. J.P. Ma is supported by National Natural Science Foundation of P.R. China(No.12075299,11821505,11935017)  and by the Strategic Priority Research Program of Chinese Academy of Sciences, Grant No. XDB34000000. X.B. Tong is supported by the CUHK-Shenzhen university development fund under grant No. UDF01001859.

\appendix

\begin{widetext}
\section{Nucleon Distribution Amplitudes}
In the appendix, we list the definitions of the nucleon distribution amplitudes used in this paper. The normalizations here follow those in \cite{Braun:1999te,Braun:2000kw}. The following notations for the quark fields or spinors are applied,
\begin{align}
q_{\uparrow(\downarrow)}=\frac{1}{2}(1\pm \gamma_5)q\ , \quad q_{\pm}= \frac{1}{2} \gamma^{\mp} \gamma^{\pm} q\ ,
\end{align} 
where the light-cone coordinates $a^{\pm}=(a^0\pm a^3)/\sqrt{2}$ are used. Suppose the nucleon state is moving along $z$ direction with the momentum $(P^+,P^-,\bs 0_\perp)$, the twist-3 nucleon distribution amplitude $\Phi_3$ is defined by
\begin{align}
&\langle  0 |\epsilon^{abc}
  \big ( \hat u^{Ta}_{+\uparrow}(z_1^- ) C \gamma^+ \hat u^b_{+ \downarrow}(z_2^-)  \big)
\hat d^c_{+\uparrow}(z_3^-) | P \rangle 
\notag \\
=& -\frac{1}{2}P^+ u_{+ \uparrow}(P) \int  [d x] \text{exp} \Big( -i  \sum^3_{i=1} x_i z_i^- P^+\Big)
\Phi_3(x_1,x_2,x_3)\ .
\label{eq_defDA1}
\end{align}
The twist-4 nucleon distribution amplitudes $\Phi_4$ and $\Psi_4$ are defined by
\begin{align}
&\langle  0 | \epsilon^{abc}
  \big( \hat u^{Ta}_{+\uparrow}(z_1^- ) C \gamma^+
\hat u^b_{+ \downarrow}(z_2^-)\big)  \gamma^- \hat d^c_{-\uparrow}(z_3^-) | P \rangle 
\notag \\
=&-\frac{M}{2} u_{+ \downarrow}(P) \int  [d x] \text{exp} \Big( -i  \sum^3_{i=1} x_i z_i^- P^+  \Big)
\Phi_4(x_1,x_2,x_3)\ ,
\notag \\
&\langle  0 | \epsilon^{abc}\big
( \hat u^{Ta}_{+\uparrow}(z_1^- ) C \gamma_\perp^i 
\hat u^b_{- \downarrow}(z_2^-) \big)\gamma_\perp^i \hat d^c_{+\downarrow}(z_3^-) | P\rangle 
\notag \\
=&\frac{M}{2} u_{+ \uparrow}(P) \int  [d x] \text{exp} \Big(  -i  \sum^3_{i=1} x_i z_i^- P^+ \Big)
\Psi_4(x_1,x_2,x_3)\ .
\label{eq_defDA2}
\end{align}
Here $\hat u, \hat d$ represent the $u,d$ quark fields respectively and the gauge links that connect the fields are implied. $C=i \gamma^2 \gamma^0$ is the charge conjugation matrix. $a,b,c$ refer to the color of the quark fields. $M$ is the nucleon mass. The label $[d x]$ denotes the measure $dx_1 dx_2 d x_3 \delta(x_1+x_2+x_3-1)$. In the Dirac representation, the Dirac spinors with definite helicity are taken as
\begin{align}
u_\uparrow(p)=
\frac{1}{\sqrt[4]{2} \sqrt{p^+}}
 \left(\begin{array}{ccc} 
p^+  \\
 (p^x+ ip^y)/\sqrt{2}  \\
p^+  \\
(p^x+ i p^y)/\sqrt{2} 
\end{array}  \right)
\ ,
\quad
u_\downarrow(p)=
\frac{1}{\sqrt[4]{2} \sqrt{p^+}}
 \left(\begin{array}{ccc} 
 - (p^x-i p^y)/\sqrt{2}\\
p^+   \\
 (p^x-i p^y)/\sqrt{2} \\
-p^+ 
\end{array}  \right)\ ,
\label{eq_spinor}
\end{align}
where the massless limit are taken.

\par 
With the solution of the the massless Dirac equation $i \slashed D\psi=0$ in the light-cone gauge $A^+=0$,
\begin{align}
 \psi_- (z^- ,z_\perp )&= -\frac{i}{2  } \int \frac{d y}{2 \pi y }  \int d \tilde  z^-  e^{i y (\tilde z^- -z^-)} \gamma^+   \slashed D_\perp \psi _+(   \tilde z^-, z_\perp)\ ,
\end{align}
and the nucleon Fock state expansion given in Eqs. (\ref{eq_fock1},\ref{eq_fock2}) as well as the definitions of the nucleon distribution amplitudes in Eqs.(\ref{eq_defDA1},\ref{eq_defDA2}), one can derive the relation between the nucleon distribution amplitudes and the light-front wave functions up to higher $g_s$ corrections:
\begin{align}
&\Phi_3 (x_1,x_2,x_3)=2\int [d^2 \bs k]\  \psi_1(\kappa_1,\kappa_2,\kappa_3)\ ,
\notag \\
&\Phi_4 (  {x_2} , {x_1},x_3)= 2 \int \frac{ [d^2 \bs k]}{M x_3}\  \bs k_{3}\cdot [ \bs  k_{1}  \psi_3(\kappa_1,\kappa_2,\kappa_3) 
+\bs  k_{2}  \psi_4(\kappa_1,\kappa_2,\kappa_3)]\ ,
\notag \\
&\Psi_4 ( x_1,x_2 ,x_3)=2\int \frac{ [d^2 \bs k]}{M x_2}\  \bs k_{2}\cdot [ \bs  k_{1}  \psi_3(\kappa_1,\kappa_2,\kappa_3) 
+\bs  k_{2}  \psi_4(\kappa_1,\kappa_2,\kappa_3)]\ ,
\end{align} 
where $\kappa_i\equiv(x_i, \bs k_i)$. The label $[d^2 \bs k]$ denotes $\frac{1}{(2\pi)^6}d^2\bs k_{1} d^2\bs k_{2} d^2\bs k_{3}\delta^{(2)}(\bs k_{1}+\bs k_{2}+\bs k_{3})$.
\section{Quark Generalized Parton Distributions at the Large $-t$}
It is well-known that the nucleon quark GPDs can be related to the nucleon quark GFFs by the following equations\cite{Ji:1996ek}
\begin{align}
\int^1_{-1} d x\  x  H_{q}(x,\xi, t)=A_q(t)+(2\xi)^2 C_q(t)~
,\quad 
\int^1_{-1} d x\  x  E_q(x,\xi, t)=B_q(t)-(2\xi)^2 C_q(t)~, 
\label{eq_GPDrelate}
\end{align}
where the relevant quark GPDs are defined by the off-forward distribution amplitude
 \begin{align}
&\int \frac{d\eta^- }{2 \pi }   e^{i x \bar P^+ \eta^- } 
\Big\langle P' ,s'\Big|
 \bar   \psi_q(-\frac{\eta^-}{2}) 
{\cal L}[-\frac{\eta^-}{2}, \frac{\eta^-}{2}]
       \psi_q(\frac{\eta^-}{2})  
  \Big| P,s \Big\rangle
   \notag \\
 =& H_q (x,\xi, t) \bar u(P',s') \gamma^+ u(P,s)
+E_q(x,\xi,t)\bar u(P',s') \frac{i \sigma^{+ \alpha} \Delta_\alpha}{2M}u(P,s)~,
 \end{align}
in which $P$ and $P'$ are the initial and final nucleon momenta respectively with $P'^2=P^2=M^2$, $s^\mu$ denote the covariant spin vector of the nucleon, and $\psi_q$ is the quark field of flavor $q$. ${\cal L}$ is the gauge link in the fundamental representation:
\begin{align}
 {\cal L} \left(z_2,z_1\right) ={\cal P}
  \ \text{exp} \left[-i g_s  \int ^{z_2}_{z_1} \  d \eta^-  G^{+,a}(\eta^- ) t^a  \right]~,
 \end{align}  
where $ t^a$ is the $SU(3)$ generator in the fundamental representation. In the definition of GPD, $\bar P=(P+P')/2$ is the average momentum, $\Delta =P'-P$ is the momentum transfer with $t=\Delta^2$. The skewness parameter $\xi$ is defined as $
\xi=\frac{ P^+-P'^+ }{P^++P'^+}$, and the light-cone coordinates $a^{\pm}=(a^0\pm a^3)/\sqrt{2}$ are used.

\par 
Similar to the GFFs, the quark GPDs can also be factorized at the large $-t$ limit. For nucleon $H_q$, it is given by the nucleon helicity-conserved amplitude and its factorization has been reported in  \cite{Hoodbhoy:2003uu}. With the relation in Eq.(\ref{eq_GPDrelate}), we have checked our result on the quark $A$-GFF is consistent with theirs. On the other hand, the quark GPD $E_q$ is related to the nucleon helicity-flip amplitude, and its behaviors at large $-t$ was unknown in the literature.  
\par   
Following the discussions in the Sec.(\ref{sec_flip}), we derive the factorization theorem for the nucleon GPD $E_q$ at large momentum transfer. The formula reads
\begin{align} 
E_q(x,\xi,t) =  \int [d x][d y]& \{ x_3 \Phi_4(x_1,x_2,x_3){\cal E}_{\Phi q}(\{x\},\{y\})
+x_1\Psi_4(x_2,x_1,x_3){\cal E}_{\Psi q}(\{x\},\{y\})
\} \Phi_3(y_1,y_2,y_3) 
+{\cal O}(t^{-4})~,
\end{align}
where the hard perturbative coefficients ${\cal E}_{\Psi q}$ and ${\cal E}_{\Phi q}$ have the following form:
\begin{align}
{\cal E}_u=2{\cal E}_1+{\cal E}_2+{\cal E}_3+{\cal E}'_1+{\cal E}'_3~,
\quad
{\cal E}_d={\cal E}_2+{\cal E}_3+{\cal E}'_2
\end{align}
with $u,d$ denote the quark flavors. 
At leading order perturbation theory, the explicit calculation yields
\begin{align}
{\cal E}_i= \frac{ C_B^2  M^2_p }{12 (-t)^3} (4\pi \alpha_s)^2 {\cal T}_i+ (\xi\leftrightarrow -\xi )~,
\end{align}
where ${\cal T}_{i\Psi}$ and ${\cal T}_{i\Phi}$ are given by
\begin{align}
{\cal T}_{1\Psi}
=&
 (1-\xi)\delta(x-\lambda_1) 
 \bigg [ 
\frac{1}{x_3 y_3^2 \bar{x}_1^2 \bar{y}_1}
 -\frac{1}{x_1 x_3 y_3 \bar{x}_1 \bar{y}_1^2}
-\frac{1}{x_1 x_2 x_3 y_2 y_3^2}
-\frac{1}{x_1 x_2  y_2 \bar{x}_1 \bar{y}_1^2}
 \bigg ]
\notag \\
&+
   (1-\xi)\delta(x-\tilde\lambda_1) 
   \bigg[ 
   \frac{1}{x_1 x_3 y_3 \bar{x}_1^2 \bar{y}_1^2}
   +\frac{1}{x_1 x_2 y_2  \bar{x}_1^2 \bar{y}_1^2}
  -  \frac{1}{x_1 x_2 x_3 y_2 y_3 \bar{x}_1 \bar{y}_3}
   \bigg] 
\notag \\
&+
(1+\xi)  \delta(x-\tilde\lambda_1)
\bigg[
\frac{1}{x_3 y_3^2 \bar{x}_1^2 \bar{y}_1}+\frac{1}{x_3 y_3 \bar{x}_1^2 \bar{y}_1^2}+\frac{1}{x_2  y_2 \bar{x}_1^2 \bar{y}_1^2}
-\frac{1}{x_2 x_3 y_2 y_3 \bar{x}_1 \bar{y}_3}
\bigg]
\notag \\ 
&+(1+\xi)\delta(x-\tilde\eta_1)\
\frac{-1}{x_1 x_2 x_3 y_2 y_3^2 \bar{y}_3}
-\frac{\delta(x-\lambda_1)- \delta(x-\tilde\eta_1) }
    {\lambda_1-\tilde \eta_1}
\frac{  -(1-\xi ^2) }{x_1 x_2 x_3 y_2 y_3^2} 
-\frac{\delta(x-\tilde \lambda_1)- \delta(x-\tilde\eta_1) }
    {\tilde\lambda_1-\tilde\eta_1}
\frac{ (1-\xi ^2) }{x_1 x_2 x_3 y_2 y_3 \bar{y}_3} 
\notag \\
&-
\frac{\delta(x-\lambda_1)- \delta(x-\tilde\lambda_1) }
    {\lambda_1-\tilde\lambda_1}
 \bigg[
  \frac{(1-\xi )^2 }{x_1 x_3 y_1 y_3 \bar{x}_1^2}  
  +\frac{ (1-\xi ^2)}{ x_3 \bar y_1  y_3^2  \bar{x}_1^2}  
   +
\frac{(1-\xi ^2)}{x_1 x_3 y_3 \bar{x}_1 \bar{y}_1^2}
+
    \frac{ (1-\xi)^2 }{x_1 \bar{x}_1^2 x_2 \bar y_1 y_2 }
+
     \frac{ (1-\xi ^2)}
     {x_1 x_2  y_2 \bar{x}_1 \bar{y}_1^2 }\bigg]
     \notag \\ 
 & +\frac{\delta(x-\lambda_1)-\delta(x-\eta_1)+\delta(x-\tilde\lambda_1)-\delta(x-\tilde\eta_1)}{(\eta_1-\tilde\lambda_1)(\lambda_1-\eta_1)}
\frac{ (1-\xi )^2 (1+\xi )}{x_1
   x_2 x_3 y_2 y_3}~,
\end{align}
\begin{align}
&\begin{aligned}
{\cal T}_{1\Phi}
=&
 (1-\xi)\delta(x-\lambda_1)
 \bigg[ \frac{1}{x_3^2 y_3^2 \bar{x}_1 \bar{y}_1}
+
\frac{1}{x_3^2 y_3 \bar{x}_1 \bar{y}_1^2}
+ \frac{1}{ x_2 x_3^2 y_2 y_3^2}
+\frac{1}{ x_2 x_3 y_2 \bar{x}_1 \bar{y}_1^2}
\bigg]
\notag \\
&+
  (1+\xi) \delta(x-\tilde\lambda_1) 
  \bigg[  \frac{1}{x_3^2 y_3 \bar{x}_1 \bar{y}_1^2}
+\frac{1}{x_3^2 y_3^2 \bar{x}_1\bar{y}_1}
+\frac{1}{x_2 x_3 y_2 \bar{x}_1 \bar{y}_1^2}
-\frac{1}{x_2 x_3^2 y_2 y_3 \bar{y}_3}
\bigg]
\notag \\ 
&+(1+\xi)\delta(x-\tilde\eta_1)\
\frac{1}{ x_2 x_3^2 y_2 y_3^2 \bar{y}_3}
-\frac{\delta(x-\lambda_1)- \delta(x-\tilde\eta_1) }
    {\lambda_1-\tilde \eta_1}
\frac{  (1-\xi ^2) }{ x_2 x_3^2 y_2 y_3^2} 
-\frac{\delta(x-\lambda_1)- \delta(x-\eta_1) }
    {\lambda_1-\eta_1}
\frac{(1-\xi)^2 }{x_2 x_3^2 y_2 y_3 \bar{x}_3}
\notag \\ 
&-
\frac{\delta(x-\tilde \lambda_1)- \delta(x-\tilde\eta_1) }
    {\tilde\lambda_1-\tilde\eta_1}
\frac{(1-\xi ^2) }{ x_2 x_3^2 y_2 y_3 \bar{y}_3} 
-
\frac{\delta(x-\lambda_1)- \delta(x-\tilde\lambda_1) }
    {\lambda_1-\tilde\lambda_1}(1-\xi ^2)
    \bigg[
\frac{ 1 }{ x_3^2 \bar y_1  y_3^2  \bar{x}_1} 
+
\frac{ 1}{ x_3^2 y_3 \bar{x}_1 \bar{y}_1^2}
    +
     \frac{  1} 
     { x_2 x_3 y_2 \bar{x}_1 \bar{y}_1^2 } \bigg]
     \notag \\ 
    &+ \frac{\delta(x-\lambda_1)-\delta(x-\eta_1)+\delta(x-\tilde\lambda_1)-\delta(x-\tilde\eta_1)}{(\eta_1-\tilde\lambda_1)(\lambda_1-\eta_1)}
\frac{ (1-\xi )^2 (1+\xi )}{
   x_2 x_3^2 y_2 y_3}~,
 \end{aligned}
\notag \\ 
&
\begin{aligned}
{\cal T}_{2\Psi}
=&
 (1-\xi) \delta(x- \lambda_2) \bigg [ 
 \frac{1}{x_1 x_3 y_1 y_3^2 \bar{x}_1}  + 
 \frac{1}{x_1^2 x_3 y_1^2 y_3 } 
+ 
 \frac{1}{x_1^2  y_1^2 \bar{x}_2 \bar{y}_2}-   \frac{1}{x_1 x_3 y_3^2 \bar{x}_2 \bar{y}_2}\bigg ]
 \notag \\ 
 &+(1+\xi)\delta(x-\tilde\lambda_2) \left [ -  \frac{1}{x_1 x_3 y_3^2 \bar{x}_2 \bar{y}_2} +  \frac{1}{x_1^2  y_1^2 \bar{x}_2 \bar{y}_2}-
\frac{1}{x_1^2 x_3 y_1 y_3 \bar{y}_1}+ 
\frac{1}{x_1^2 x_3 y_1 y_3 \bar{y}_3} \right]
\notag \\ 
&+
(1+\xi)\delta(x-\eta_2)  \frac{1}{x_1^2 x_3 y_1^2 y_3 \bar{y}_1}
+(1+\xi)\delta(x-\tilde\eta_2) \frac{1}{x_1 x_3 y_1 y_3^2 \bar{x}_1 \bar{y}_3}
+(1-\xi) \delta(x-\tilde\eta_2)  \frac{1}{x_1^2 x_3 y_1 y_3 \bar{x}_1 \bar{y}_3}   
\notag \\ 
&-(1+\xi)\delta(x-\tilde\eta_1)
\frac{1}{x_1 x_2 x_3 y_2 y_3^2 \bar{y}_3}
+
(1+\xi)\delta(x-\tilde\eta_3)
\frac{1}{ x_2 x_1^2 y_2 y_1^2 \bar{y}_1}
\notag \\ 
&-\frac{\delta(x-\lambda_2)- \delta(x- \eta_3) }
    {\lambda_2-\eta_3} \left [ 
\frac{ (1-\xi )^2}{x_1^2 x_3 y_1 y_3 \bar{x}_1}
+\frac{ (1-\xi ^2)}{x_1 x_3 y_1 y_3^2 \bar{x}_1} \right]
-\frac{\delta(x-\tilde\lambda_2)- \delta(x-\tilde\eta_3) }
    {\tilde\lambda_2-\tilde\eta_3}
\frac{  (1-\xi ^2)
  }{x_1^2 x_3 y_1  \bar{y}_1 y_3}
\notag \\ &
-
\frac{\delta(x-\lambda_2)- \delta(x- \eta_2) }
    {\lambda_2-\eta_2}
\frac{(1-\xi ^2)}{x_1^2 x_3 y_1^2 y_3 } 
-
\frac{\delta(x-\tilde\lambda_2)- \delta(x-\tilde\eta_2) }
    {\tilde\lambda_2-\tilde\eta_2}
\frac{ (1-\xi ^2) }{x_1^2 x_3 y_1 y_3 \bar{y}_3} 
\notag \\ &
-\frac{\delta(x-\lambda_2)- \delta(x-\tilde\lambda_2) }
    {\lambda_2-\tilde\lambda_2}\left(1-\xi ^2\right)
\bigg [  \frac{ -1 }{x_1 \bar x_2 x_3 \bar  y_2 y_3^2}
 + \frac{ 1
    }{x_1^2 \bar x_2  y_1^2 \bar y_2}\bigg  ]
   \notag \\ &
 + \frac{\delta(x-\lambda_2)-\delta(x-\eta_2)+\delta(x-\tilde\lambda_2)-\delta(x-\tilde\eta_2)}{(\eta_2-\tilde\lambda_2)(\lambda_2-\eta_2)}\frac{  (1-\xi )^2 (1+\xi )}{x_1^2 x_3 y_1 y_3}~,
 \end{aligned}
  \end{align}
 and   ${\cal T}_{3\Phi}={\cal T}_{1\Psi}(1\leftrightarrow3)$, ${\cal T}_{3\Psi} ={\cal T}_{1\Phi}(1\leftrightarrow3)
$, and $ {\cal T}_{2\Phi} ={\cal T}_{2\Psi}(1\leftrightarrow3)
$, where $\lambda_i =y_i+ \bar y_i \xi$,\ $
\tilde \lambda_i=x_i- \bar x_i\xi$ and 
\begin{align}
&\eta_1=1-x_3-y_2+(y_2-x_3)\xi~,\quad
\tilde \eta_1=1-y_3-x_2-(x_2-y_3)\xi~,
\notag \\
& \eta_2=1-x_3-y_1+(y_1-x_3)\xi
~, \quad 
\tilde \eta_2
=1-y_3-x_1-(x_1-y_3)\xi
~,
\notag \\
&\eta_3=1-x_1-y_3+(y_3-x_1)\xi~,\quad
\tilde \eta_3=1-x_3-y_1-(y_3-x_1)\xi~. 
\end{align}

\end{widetext}

%

\end{document}